\documentclass[11pt,a4paper]{article}
\pdfoutput=1
\usepackage{myjheppub}
\usepackage{latexsym,amsfonts,amsmath,amssymb,dsfont} 
\usepackage{amsthm}
\usepackage{mathtools}
\usepackage{bbm}
\usepackage[abs]{overpic}
\usepackage{graphicx}
\usepackage[dvipsnames]{xcolor}
\usepackage{caption}
\usepackage{subcaption}      
\usepackage[normalem]{ulem}
\usepackage{comment}
\usepackage{url}
\usepackage{slashed}
\usepackage{cancel}
\usepackage{bm}
\usepackage{hhline}
\usepackage{tabu}
\usepackage{physics}

\usepackage[colorlinks=true
,urlcolor=blue
,anchorcolor=blue
,citecolor=blue
,filecolor=blue
,linkcolor=blue
,menucolor=blue
,linktocpage=true
]{hyperref}

\renewcommand{\tilde}{\widetilde}

\renewcommand{\Re}{\operatorname{Re}}
\renewcommand{\Im}{\operatorname{Im}}

\DeclareMathOperator{\Arg}{arg} 
\DeclareMathAlphabet{\mathbfsf}{OT1}{cmss}{bx}{n}

\renewcommand{\abs}[1]{\left\lvert #1 \right\rvert}

\newcommand{\Z}{\mathbb{Z}}

\newcommand{\R}{\mathbb{R}}

\newcommand{\mc}[1]{\mathcal{#1}}

\newcommand{\cA}{\mathcal{A}}

\newcommand{\cF}{\mathcal{F}}

\newcommand{\cH}{\mathcal{H}}
\newcommand{\cI}{\mathcal{I}}

\newcommand{\cN}{\mathcal{N}}

\newcommand{\cQ}{\mathcal Q}

\newcommand{\cU}{\mathcal{U}}

\newcommand{\be}{\begin{equation}}
\newcommand{\ee}{\end{equation}}

\newcommand{\beq}{\begin{equation}}
\newcommand{\eeq}{\end{equation}}

\newcommand{\ii}{{\rm i}}
\newcommand{\e}{{\rm e}}

\newcommand{\rd}{{\rm d}}

\newcommand{\vol}{{\rm vol}}
\newcommand{\ph}[1]{\phantom{#1}}

\renewcommand{\=}{\, = \,}

\allowdisplaybreaks 

\DeclareFontShape{OT1}{cmr}{mx}{n}%
{<->cmr10}{}

\DeclareMathAlphabet{\titlemath}{OT1}{cmr}{mx}{n}

\newcommand{\tE}{t_E}

\newcommand{\phiCanonicalPeriodicities}{\phi}
\newcommand{\BulkTheta}{\theta}
\newcommand{\BdryTheta}{\vartheta}

\newcommand{\psiCanonicalPeriodicities}{\psi}

\newcommand{\Periodictau}{t_\text{E}}

\newcommand{\FirstParameterSCI}{\sigma}
\newcommand{\SecondParameterSCI}{\tau}

\newcommand{\EnergyOp}{E}

\title{Allowable complex metrics and the gravitational index of AdS$_5$ black holes}

\author{Pietro Benetti Genolini,${}^1$ Oliver Janssen${}^2$ and Sameer Murthy${}^{3}$}

\affiliation{${}^1$ D\'epartement de Physique Théorique, 
Universit\'e de Gen\`eve, 24 quai Ernest-Ansermet, \\ $\phantom{f}$ 1211 Gen\`eve, Suisse}
\affiliation{${}^2$ Laboratory for Theoretical Fundamental Physics, EPFL, 1015 Lausanne, Switzerland}
\affiliation{${}^3$ Department of Mathematics, King's College London, The Strand, London WC2R 2LS, UK}

\emailAdd{pietro.benettigenolini@unige.ch, oliver.janssen@epfl.ch, sameer.murthy@kcl.ac.uk}

\abstract{
We discuss the Kontsevich--Segal--Witten criterion for the allowability of complex metrics, in the 
context of the gravitational path integral that calculates the supersymmetric index. 
We focus on the saddle points 
that capture the contribution of supersymmetric black holes in AdS$_5$ space. 
We show that, for such black holes with two independent angular momenta, 
the conditions imposed on the corresponding saddle point by the KSW criterion are   
equivalent to the ones arising from the convergence of the microscopic trace form of the supersymmetric index.
This result adds to previous results establishing such an equivalence in other, simpler examples of the gravitational index in AdS space and flat space.
Along the way, we give a practical algorithm for implementing the KSW criterion in terms of eigenvalues of certain matrices.
}

\begin{document}

\maketitle

\section{Introduction \label{sec:Intro}}

In classical General Relativity, the metric on a manifold is a symmetric second-rank tensor that takes real values at every point. 
However, there are many situations in physics where complex-valued metrics naturally arise, especially in the context of the Gravitational Path Integral (GPI) as discussed by Gibbons and Hawking~\cite{Gibbons:1976ue}. 
A simple example is the analytic continuation to Euclidean signature (the ``Wick rotation") of a rotating black hole solution, 
which solves the saddle point equations of the GPI.
It is clear that not all complex metrics are physical, and 
extending the space of integration of the GPI to all complex metrics is not sensible.  
This prompts the question of which complex metrics are allowed as saddle points of the GPI. A criterion to answer this question was proposed by Witten~\cite{Witten:2021nzp}, based on the work of Kontsevich and Segal~\cite{Kontsevich:2021dmb} (henceforth referred to as the KSW criterion).

In order to check the validity of such a criterion, it is useful to consider situations in which we have independent knowledge and control over the GPI.
One such context is the gravitational supersymmetric index
(see \cite{Cassani:2025sim} for a review of the concept).\footnote{This is by no means the only context where the KSW criterion has been deployed, see e.g.~\cite{Lehners:2021mah, Visser:2021ucg, Rajeev:2021xit, Chen:2022hbi, Muhlmann:2022duj, Loges:2022nuw, Narain:2022msz, Jonas:2022uqb, Briscese:2022evf, Rosso:2022tsv, Lehners:2022xds, Usatyuk:2022afj, Bobev:2022lcc, Bah:2022uyz, Basile:2023ycy, Chen:2023prz, Hertog:2023vot, Chen:2023sry, Chen:2023mbc, Chen:2023hra, Chen:2024vpa, Maldacena:2024uhs, Grabovsky:2024vnb, Janssen:2024vjn, Wen:2024bzm, Chakravarty:2024bna, Hertog:2024nbh, Fumagalli:2024msi, Ivo:2024ill, Sivakumar:2024iqs, Matone:2024ytm, Hertog:2024shf, Chakravarty:2025sbg, Horowitz:2025zpx, Ailiga:2025fny, Hull:2025rxy, DHoker:2025nid, Mahajan:2025bzo, Jones:2025gno, Ailiga:2025osa} for a sample of the ever-growing list of applications.}
The index is protected against changes of coupling constant, and therefore it can be directly compared with a corresponding calculation of the index in microscopic string theory. 
In cases that allow for such a calculation, independent ideas and constraints from the microscopic theory, such as 
modular symmetry and saddle point solutions of matrix models,
predict a series of complex saddles of the GPI. 

\smallskip

In~\cite{BenettiGenolini:2025jwe, MasterThesis}, an exploration of whether complex saddles of the gravitational index are allowed by the KSW criterion was initiated. 
Different types of complex black-hole-like saddles were considered for the validity of the KSW criterion as well as the microscopic constraints. 
The result was that---by and large---complex black-hole-like saddles of the gravitational index are allowed by the KSW criterion exactly when they are allowed by the microscopic constraints. 
However, the case of supersymmetric black holes in AdS$_5$ space carrying two unequal angular momenta was puzzling, and was not satisfactorily treated. 
As it turns out, this puzzle stems from an error in the implementation of the criterion (which only affects the case of different angular momenta) in a first version of~\cite{BenettiGenolini:2025jwe}. 
As we explain in the present paper, the proper implementation of the KSW criterion shows that the KSW criterion agrees precisely with the microscopic constraints for the asymptotically AdS$_5$ black hole.\footnote{We have also updated~\cite{BenettiGenolini:2025jwe} with a new version. The discussion of the $5d$ GPI dual to the superconformal index in that paper is superseded by the present paper and hence has been removed from there.}
Upon combining this with the other results of~\cite{BenettiGenolini:2025jwe}, we reach the statement that in the variety of saddles of the gravitational index that we have checked, 
namely $4d$ asymptotically flat, $4d$ asymptotically AdS with twisted and untwisted supersymmetry-preserving boundary conditions, 
and also $5d$ asymptotically AdS gravity,
there is a complete agreement of the microscopic constraints with the KSW criterion.

Given the indications that such an agreement may not hold in more generality for the GPI (see e.g.~\cite{Maldacena:2019cbz, Mahajan:2021nsd, Bah:2022uyz, Chen:2023hra}),
it would be interesting to explain why the KSW criterion works so well for the index.
If it is true as a fact, then it may be explainable using the existence of a Killing spinor in such geometries, but we do not have such an explanation so far. 
Since a sizable set of examples of gravitational indices have been found in recent years in AdS space~\cite{Hosseini:2017mds,Hosseini:2018dob,Cabo-Bizet:2018ehj,Choi:2018fdc,Cassani:2019mms,Kantor:2019lfo,Bobev:2019zmz,Bobev:2020pjk,Larsen:2021wnu,Cassani:2021dwa,BenettiGenolini:2023rkq, BenettiGenolini:2023ucp, Bobev:2023bxl, BenettiGenolini:2024lbj, Bobev:2025xan}
as well as in flat space~\cite{Iliesiu:2021are,Hristov:2022pmo,H:2023qko,Boruch:2023gfn,Cassani:2024kjn,Boruch:2025qdq,Boruch:2025biv, Boruch:2025sie}, 
one could try to find patterns in these examples in order to probe this idea.

\smallskip

In the process of implementing the KSW criterion for the AdS$_5$ black hole, we also gain a better understanding of the criterion in holographic situations, i.e.~when we have asymptotically AdS boundary conditions on the metric. 
The summary of the picture that we develop, as explained in Section~\ref{sec:KSWAdS5}, is the following. 
Recall that the KSW criterion is a point-wise criterion that must be satisfied at all points in the space. 
Applied to the asymptotic region, it is essentially the same as the original criterion of~\cite{Kontsevich:2021dmb} 
for coupling the boundary CFT to gravity.
For the supersymmetric index that we consider in this paper, 
this condition is equivalent to the convergence of the microscopic trace on BPS states.
Once we have established the equivalence at the boundary, we want to ``pull it into the bulk." 
Using the Fefferman--Graham expansion we can do so analytically in perturbation theory, but this does not really shed light on the interior.
What we do instead is to analyze the deep interior, close to the horizon, 
by imposing reality conditions justified by the physical picture. 
In the space between the asymptotic boundary and the horizon, we do not have an analytical understanding and we resort to numerical calculations, which show
that there is a weakening of the KSW condition as one moves into the interior from the boundary, i.e., 
the area in parameter space ruled out by the KSW criterion becomes smaller as we move into the interior.
Unfortunately, we were not able to capture this area as a simple analytic function of the metric. 
It would be nice to have a conceptual explanation of this weakening. 

\smallskip

Before outlining the structure of the paper, we mention a few natural generalizations of our results. 
Firstly, here we only consider non-extremal deformations of supersymmetric black holes in minimal gauged supergravity, which contribute to the gravitational supersymmetric index without refinements due to global flavor symmetry groups. 
Supersymmetric non-extremal deformations of charged rotating black holes are known in $5d$  $U(1)^3$ gauged 
supergravity~\cite{Cassani:2019mms}, and one could also test 
the KSW criterion in those situations.

Secondly, there are also Abelian gauge fields in the supersymmetric solutions considered that have complex values, 
and we should try to expand the criterion to include such backgrounds.
One possibility would be to uplift the theory to higher dimensions, in which case the 
5$d$ gauge field is geometrized to the higher-dimensional metric. 
The solutions considered here have been uplifted on Sasaki--Einstein 5-manifolds to solutions of 10-dimensional Type IIB theory~\cite{Buchel:2006gb, Gauntlett:2007ma}. However, in these constructions, the RR-form is complex as well as the metric, and so the problem of a complex gauge field persists.

Finally, we note that the authors of~\cite{Aharony:2021zkr} proposed a criterion for the inclusion of saddle points in the GPI in AdS$_5 \times S^5$ dual to $4d$ $\cN=4$ SYM. 
Their criterion is that a solution should be included in the GPI only if the action of any D3-brane wrapping a maximal~$S^3$ in~$S^5$ and an $S^1$ in the $S^3$ horizon of the AAdS$_5$ black hole satisfies $\Im (S_{\rm D3}) > 0$. 
For the solutions of minimal gauged supergravity considered in this paper, this D-brane stability criterion is satisfied in a region in parameter space containing the one defined by the KSW criterion, namely, the one defined by the geometric constraints discussed in Section \ref{subsec:ComplexDeformations} \cite{Aharony:2021zkr,Aharony:2024ntg}. 
In order to explore the relation between the D-brane stability criterion and the KSW criterion, we would need explicit constructions of supersymmetric solutions with chemical potentials obtained from those in the current paper by integer shifts.

\smallskip

The rest of this paper is organized as follows. In Section~\ref{sec:susyBHAdS5} we review the supersymmetric black hole in AdS$_5$ space and its complex continuation. In Section~\ref{sec:KSWAdS5} we discuss the application of the KSW criterion to this complex metric. 
Apart from probing the theoretical ideas behind the KSW criterion, we also develop an algorithm 
that allows us to work with eigenvalues rather than with the quadratic form that is intrinsic to the criterion. 
This algorithm is useful for numerically ruling out a given complex metric according to KSW. We explain this in Appendix~\ref{appA}.

\section{Supersymmetric black holes in \texorpdfstring{AdS$_5$}{AdS5} and complex metrics \label{sec:susyBHAdS5}}


\subsection{Lorentzian supersymmetric black holes}

Here, we review the gravity solution we work with, as introduced by \cite{Cabo-Bizet:2018ehj}, first describing its Lorentzian origin. The complex metric depends on three real parameters $(a, r_+, r_\star)$, and we describe the constraints among them that are imposed by the well-definedness of the original Lorentzian metric, and by the convergence of the microscopic description of the dual superconformal index.

\smallskip

We focus on minimal gauged supergravity in five dimensions, whose bosonic sector includes a metric and an Abelian gauge field $\cA$ with field strength $\cF = \rd \cA$, interacting via the action
\begin{equation}
\label{eq:AdS_5d_Theory}
    S \= \frac{1}{16\pi} \int \left[ \left( R + 12 - \frac{1}{3} \cF^2 \right) \vol + \frac{8}{27} \cA \wedge \cF \wedge \cF \right] \, .
\end{equation}
This theory admits an AdS$_5$ solution with unit radius, and a family of solutions that describe rotating, electrically charged black holes \cite{Chong:2005hr}
\begin{align}
\label{eq:AdS_KN_5d_Metric_App}
\begin{split}
    \rd s^2 &\= - \frac{\Delta_\BulkTheta [(1 + r^2 )\rho^2 \, \rd t + 2 q \nu] \rd t}{\Xi_a \Xi_b \rho^2} + \frac{2q \nu\omega}{\rho^2} + \frac{f}{\rho^4} \left( \frac{\Delta_\BulkTheta}{\Xi_a \Xi_b} \rd t - \omega \right)^2 \\
    & \ \ \ \ \, + \frac{r^2 + a^2}{\Xi_a} \sin^2\BulkTheta \left( \rd\phiCanonicalPeriodicities + \Omega_1 \, \rd t \right)^2 + \frac{r^2 + b^2}{\Xi_b} \cos^2\BulkTheta \left( \rd\psiCanonicalPeriodicities + \Omega_2 \,\rd t \right)^2 \\
    & \ \ \ \ \, + \rho^2 \left( \frac{\rd r^2}{\Delta_r} + \frac{\rd\BulkTheta^2}{\Delta_\BulkTheta} \right) \,, 
\end{split}     \\[10pt]
\label{eq:AdS_KN_5d_GaugeField}
    \cA &\= \frac{3 q}{2\rho^2 } \left( \frac{\Delta_\BulkTheta}{\Xi_a \Xi_b} \rd t - \omega \right) - \frac{3q r_+^2}{2\left[ (r_+^2 + a^2)(r_+^2+b^2)+abq \right]} \, \rd t \, ,
\end{align}
where
\begin{equation}
\label{eq:AdS_KN_5d_Definitions}
\begin{split}
    \nu &\= b \sin^2 \BulkTheta \left( \rd\phiCanonicalPeriodicities + \Omega_1 \, \rd t \right) + a \cos^2\BulkTheta \left( \rd\psiCanonicalPeriodicities + \Omega_2 \,\rd t \right) \, , \\ 
    \omega &\= \frac{a \sin^2\BulkTheta}{\Xi_a} \left( \rd\phiCanonicalPeriodicities + \Omega_1 \, \rd t \right) + \frac{b \cos^2\BulkTheta}{\Xi_b} \left( \rd\psiCanonicalPeriodicities + \Omega_2 \,\rd t \right) \, , \\
    \Delta_r &\= \frac{(r^2 + a^2)(r^2 + b^2)(1 + r^2) + q^2 + 2abq}{r^2} - 2m \, , \\
    \Delta_\BulkTheta &\= 1 - a^2 \cos^2\BulkTheta - b^2 \sin^2\BulkTheta \, , \qquad \quad \quad \rho^2 \= r^2 + a^2 \cos^2\BulkTheta + b^2 \sin^2\BulkTheta \, , \\
    \Xi_a &\= 1 - a^2 \, , \qquad \Xi_b \= 1 - b^2 \, , \qquad \qquad f \= 2m\rho^2 - q^2 + 2abq \rho^2 \, , \\
    \Omega_1 &\= \frac{ a(r_+^2 + b^2)(1 + r_+^2) + bq }{(r_+^2 + a^2)(r_+^2 + b^2) + ab q} \, , \qquad \ \;  \Omega_2 \= \frac{ b(r_+^2 + a^2)(1 + r_+^2) + aq }{(r_+^2 + a^2)(r_+^2 + b^2) + ab q} \,.
\end{split}
\end{equation}
The coordinate $r$ is larger than $r_+$, the largest positive root of $\Delta_r$, and $\phiCanonicalPeriodicities$ and $\psiCanonicalPeriodicities$ are periodic with period $2\pi$, describing a three-sphere as a torus fibration over an interval parametrized by $\BulkTheta\in [ 0,\pi/2]$. The gauge field \eqref{eq:AdS_KN_5d_GaugeField} is written in the ``regular'' gauge, where the one-form is regular on the Wick-rotated geometry.
These solutions depend on four parameters $(m,q,a,b)$, with
\begin{equation}
\label{eq:Geometric_Constraint_Degeneracy_v1}
    a^2 \, < \, 1 \, , \qquad b^2 \, < \, 1 \,,
\end{equation}
in order to avoid degenerate points for the metric (so $\Xi_{a,b}>0$), and describe black holes with the following energy, charge and angular momenta:
\begin{equation}
\begin{aligned}
    E &\= \frac{\pi}{4 \Xi_a^2 \Xi_b^2} \left( m (2\Xi_a + 2\Xi_b - \Xi_a \Xi_b) + 2 q ab (\Xi_a + \Xi_b) \right) \, , \\ 
    Q &\= \frac{\pi q }{2 \Xi_a \Xi_b} \,, \\
    J_1 &\= \frac{\pi}{4 \Xi_a^2 \Xi_b} \left( 2am + qb (1 + a^2) \right) \,, \\
    J_2 &\= \frac{\pi}{4 \Xi_a \Xi_b^2} \left( 2bm + qa (1 + b^2) \right) \, .
\end{aligned}
\end{equation}
The event horizon at $r=r_+$ is described by its temperature, whose inverse is
\begin{equation}
    \beta \= 2\pi \frac{r_+ [(r_+^2 + a^2)(r_+^2 + b^2) + abq]}{r_+^4 [ 1 + (2r_+^2 + a^2 + b^2)] - (ab + q)^2} \, ,
\end{equation}
the angular velocities are $\Omega_1$ and $\Omega_2$ in \eqref{eq:AdS_KN_5d_Definitions}, the electrostatic potential reads
\begin{equation}
    \Phi_e \= - \frac{3q r_+^2}{2\left( (r_+^2 + a^2)(r_+^2+b^2)+abq \right)} \, ,
\end{equation}
and the associated entropy is
\begin{equation}
    S \= \pi^2 \frac{(r_+^2 + a^2)(r_+^2 + b^2) + abq}{2 \Xi_a \Xi_b r_+} \, .
\end{equation}
The Killing generator of the horizon is $V = \partial_t$.

\subsection{Complex deformations via analytic continuation}
\label{subsec:ComplexDeformations}

A supersymmetric solution of the theory \eqref{eq:AdS_5d_Theory} supports a global Dirac spinor $\epsilon$ satisfying
\begin{equation}
\label{eq:AdS_5d_KSE}
    \left( \nabla_\mu - \ii \cA_\mu - \frac{1}{2} \Gamma_\mu - \frac{\ii}{12} \left( \Gamma_\mu^{\ph{\mu}\nu\rho} - 4 \delta_\mu^\nu \Gamma^\rho \right) \cF_{\nu\rho} \right) \epsilon \= 0 \, ,
\end{equation}
where $\Gamma_\mu$ generate Cliff$(1,4)$. Among the family of solutions \eqref{eq:AdS_KN_5d_Metric_App} sits the two-parameter family of supersymmetric, extremal, rotating, electrically charged black holes with two unequal angular momenta \cite{Chong:2005hr}, which is found by imposing\footnote{Further setting $a=b$ gives the black hole found by Gutowski and Reall \cite{Gutowski:2004ez}.}
\begin{equation}
\label{eq:5d_BPS_Constraints}
    q \= \frac{m}{1+a+b} \, , \qquad m \= (a+ b)(1 + a )( 1 + b )( 1 + a+ b) \, , 
\end{equation}
in which case $\Delta_r$ becomes
\begin{equation}
    \Delta_r \= \frac{\left(r^2 - (ab + a + b ) \right)^2 \left((1 + a + b)^2+r^2\right)}{r^2 } \, .
\end{equation}
We find a double root for $\Delta_r$, signalling extremality, provided
\begin{equation}
\label{eq:Geometric_Constraint_rstarReal}
    ab + a + b \, > \, 0 \, ,
\end{equation}
in which case the extremal horizon sits at
\begin{equation}
\label{eq:Defn_rstar}
    r_\star \= \sqrt{ab + a + b} \, ,
\end{equation}
which can be inverted to give $b$ in terms of $r_\star^2$ (as will be useful later on),
\begin{equation}
\label{eq:Defn_b}
    b \= \frac{r_\star^2 - a }{1+a} \, .
\end{equation}
The supersymmetric black hole (necessarily extremal) is characterized by
\begin{equation}
\label{eq:BPS_Physics_Charges}
\begin{aligned}
    E_\star &\= \frac{\pi  \left[ 3 \left( a(1-a + r_\star^2) + 1 \right) - r_\star^4 -a^2 r_\star^2 \right] \left(a^2+r_\star^2\right)}{4 (1-a)^2 \left( 2 a + 1 - r_\star^2 \right)^2} \, , \\
    Q_{\star} &\= \frac{\pi  \left(a^2+r_\star^2\right)}{2 (1-a) \left( 2 a + 1 - r_\star^2 \right)} \, , \\
    J_{1\star} &\= \frac{\pi  \left(a+r_\star^2\right) \left(a^2+r_\star^2\right)}{4 (1-a)^2 \left( 2 a + 1 - r_\star^2 \right)} \, , \\
    J_{2\star} &\= \frac{\pi  \left(a^2+r_\star^2\right) \left((a+2) r_\star^2-a\right)}{4 (1-a) \left( 2 a + 1 - r_\star^2 \right)^2} \, , &\quad 
\end{aligned}
\end{equation}
which are related by 
\begin{equation}
\label{eq:BPS_Bound}
    E_\star \= J_{1\star} + J_{2\star} + \frac{3}{2} Q_\star \, .
\end{equation}
Moreover, we have
\begin{equation}
\label{eq:BPS_Physics_Potentials}
\begin{aligned}
    \beta_\star &\= \infty \, , \qquad \Phi_{e\star} \= \frac{3}{2} \, , \qquad \Omega_{1\star} \= \Omega_{2\star} \= 1 \, , \\
    S_\star &\= \pi^2 \frac{ r_\star \left(a^2+r_\star^2\right)}{2 (1-a) \left(2 a + 1 - r_\star^2 \right)} \= \pi \sqrt{3Q_\star^2 - \pi (J_{1\star} + J_{2\star} )} \, .
\end{aligned}
\end{equation}

\smallskip

We are interested in a supersymmetry-preserving complex deformation of the above solution, first introduced in \cite{Cabo-Bizet:2018ehj}. We first perform a Wick rotation $t= - \ii \beta \tE$ in \eqref{eq:AdS_KN_5d_Metric_App} and \eqref{eq:AdS_KN_5d_GaugeField}, obtaining a complex metric tensor and gauge field labelled by $(m, q, a, b)$ and defined on $\R^2\times S^3$, provided $\tE\sim \tE + 1$, and the angular coordinates maintain the periodicities already introduced below \eqref{eq:AdS_KN_5d_Definitions}. We then impose the first constraint in \eqref{eq:5d_BPS_Constraints}
\begin{equation}
\label{eq:AdS5_mSUSY}
    m \= q ( 1 + a + b) \, , 
\end{equation}
which guarantees the existence of a Dirac spinor satisfying the analytic continuation of \eqref{eq:AdS_5d_KSE}, and that the conserved charges of the spacetime satisfy the supersymmetry relation \eqref{eq:BPS_Bound} even though the analytic continuation of $\beta$ doesn't diverge\footnote{More generally, supersymmetric solutions of the Euclideanized theory have complex metric and gauge field, but they also support two independent Dirac spinors, due to the doubling of degrees of freedom. Here we don't consider this case, though see \cite{BenettiGenolini:2025icr} for details.}
\begin{equation}
\label{eq:5d_AdSKN_BPSBound}
    \EnergyOp \= J_1 + J_2 + \frac{3}{2}Q_e \, .
\end{equation}
The resulting metric and gauge field are labelled by $(q,a,b)$, though it is more useful to trade $b$ for $r_\star$ using \eqref{eq:Defn_b}, and exchange $q$ for $r_+$, using its definition as the largest positive root of $\Delta_r$, which we assume to satisfy $r_+ > r_\star$:
\begin{equation}
\label{eq:AdS5_qSUSY}
\begin{split}
    q &\= - ( a \pm \ii r_+)(b \pm \ii r_+)(1 \pm \ii r_+) \\
    &\= \frac{ a^2 (1 + r_+^2) + r_+^2 (1 + r_\star^2) + ( a \pm \ii r_+ a \pm \ii r_+) (r_+^2 - r_\star^2) }{1+a} \, .
\end{split}
\end{equation}
Therefore, a supersymmetric solution is described by three numbers $(a,r_+,r_\star)$, which we take to be all real (in contrast, $q$ and $m$ fixed via \eqref{eq:AdS5_qSUSY} and \eqref{eq:AdS5_mSUSY} are complex and $b$ fixed via \eqref{eq:Defn_b} is real).
However, they cannot take any value: the regularity condition \eqref{eq:Geometric_Constraint_Degeneracy_v1} translates to
\begin{equation}
\label{eq:Geometric_Constraint_Degeneracy_v2}
    a^2 \, < \, 1 \, , \quad 
    b^2 \, < \, 1 \qquad 
    \Longleftrightarrow \qquad 
    \frac{r_\star^2 - 1}{2} \, < \, a \, < \, 1 \,, \qquad 
    0 \,< \, r_\star^2 \, < \, 3 \, .
\end{equation}
Requiring that $r_\star\in\R$ is equivalent to \eqref{eq:Geometric_Constraint_rstarReal}, and we further assume that 
\begin{equation}
\label{eq:Geometric_Constraint_rPlus_rStar}
    r_+ \, > \, r_\star \, > \, 0 \, .
\end{equation}
The thermodynamic potentials of the complex supersymmetric solutions are
\begin{equation}
\label{eq:SUSY_Thermodynamic_Potentials}
\begin{split}
    \beta &\= 2\pi \frac{(a \pm  \ii r_+)(b \pm \ii r_+)\left( r_\star^2 \mp i r_+ \right)}{(r_+^2 -r_\star^2)\left[2 (1+a+b) r_+  \mp \ii (r_\star^2 - 3 r_+^2)\right]} \, , \\
    \Omega_1 &\= \frac{(r_+ \mp \ii ) \left(r_\star^2 \mp \ii a r_+ \right)}{ \left( r_\star^2 \mp \ii  r_+ \right) (r_+ \mp \ii a ) } \ ,\qquad
    \Omega_2 \= \frac{(r_+ \mp \ii ) \left(r_\star^2 \mp \ii b r_+  \right)}{ \left( r_\star^2 \mp \ii  r_+ \right) (r_+ \mp \ii b ) } \ , \\[1mm]
    \Phi_e &\=  \frac{3}{2} \frac{(r_+^2 \mp \ii r_+ )}{(r_\star^2 \mp \ii r_+ )} \ ,
\end{split}
\end{equation}
where $b$ is determined by \eqref{eq:Defn_b}, and the signs refer to the choice of branch in the square root in \eqref{eq:AdS5_qSUSY}. They satisfy 
\begin{equation}
\label{eq:5d_AdSKN_ChemicalPotentials_Constraint}
    \frac{\beta}{2\pi\ii} ( 1 - 2 \Phi_e + \Omega_1 + \Omega_2) \= \mp 1 \, , 
\end{equation}
which guarantees that the supersymmetric spinor $\epsilon$ is anti-periodic when parallel-transported around $\partial_{t_E}$ (in the regular gauge for the gauge field). We also notice that the conditions \eqref{eq:Geometric_Constraint_Degeneracy_v2} and \eqref{eq:Geometric_Constraint_rPlus_rStar} imply the absence of velocity-of-light surfaces~\cite{Hawking:1999dp}, i.e., 
\begin{equation}
    \abs{\Omega_1} \, < \, 1 \, , \qquad \abs{\Omega_2} \, < \, 1 \, ,
\end{equation}
with the values of the potentials given in \eqref{eq:SUSY_Thermodynamic_Potentials}.

\subsection{Boundary behavior and match with the superconformal index}

In order to describe the role of these supersymmetric complex deformations in the context of the \mbox{AdS/CFT} correspondence, we study their boundary behaviour, for which it is useful to perform the following coordinate change $(r,\theta) \to (z,\vartheta)$~\cite{Hawking:1998kw},
\begin{equation}
\label{eq:AdS5_BdryCoords}
    \frac{\Xi_a \sin^2\BdryTheta}{z^2} \= (r^2 + a^2) \sin^2\BulkTheta \, , \qquad \frac{\Xi_b \cos^2\BdryTheta}{z^2} \= (r^2 + b^2) \cos^2\BulkTheta \, .
\end{equation}
At leading order as~$r \to \infty$ or, equivalently,~$z \to 0$, we have 
\begin{align}
\label{eq:AdS5_BdryMetric}
    \rd s^2 &\, \sim \, \frac{\rd z^2}{z^2} + \frac{1}{z^2} \Bigl( \beta^2 \rd \Periodictau^2 + \rd\BdryTheta^2 + \sin^2\BdryTheta \, 
    (\rd\phiCanonicalPeriodicities - \ii \beta\Omega_1 \, \rd \Periodictau)^2 + \cos^2\BdryTheta \, (\rd\psiCanonicalPeriodicities - \ii \beta\Omega_2 \, \rd \Periodictau)^2 \Bigr) \, , \nonumber \\
    \cA & \, \sim \,   \ii \beta\Phi_e \, \rd\Periodictau \, ,
\end{align}
which describe the conformal class of backgrounds where the dual SCFT$_4$ would be formulated.

Indeed, a four-dimensional~$\cN=1$ SCFT$_4$ with a $U(1)_R$ R-symmetry can be formulated on~$S^1\times S^3$ preserving two supercharges by choosing the following configuration of metric and R-symmetry gauge field  (for concreteness, we take the~$S^1$ to have circumference~$\beta$ and $S^3$ to have unit radius)
\begin{equation}
\label{eq:4d_SCI_Background}
\begin{split}
    \rd s^2_{\rm bdry} &\= \beta^2\rd\Periodictau^2 +  \rd\BdryTheta^2 + \sin^2\BdryTheta \bigl( \rd \phiCanonicalPeriodicities - \ii \beta\Omega_1 \, \rd\Periodictau \bigr)^2 
    + \cos^2\BdryTheta \bigl( \rd\psiCanonicalPeriodicities - \ii \beta\Omega_2 \, \rd\Periodictau \bigr)^2 \,, \\
    A_R &\= \ii \beta\Phi_R \, \rd\Periodictau \, ,
\end{split}
\end{equation}
where $\Periodictau \sim \Periodictau + 1$ is the coordinate on the circle, and (as before) we have written the~$S^3$ as a torus fibration over the interval, 
so that~$\BdryTheta \in[ 0,\pi/2]$, $\phiCanonicalPeriodicities \sim \phiCanonicalPeriodicities + 2\pi$, $\psiCanonicalPeriodicities \sim \psiCanonicalPeriodicities + 2\pi$.

The states obtained by quantizing the theory on $S^3$ are labelled by the quantum numbers~$\{\EnergyOp, J_1, J_2, R\}$ corresponding to the energy $E$, the generators $J_1, J_2$ of the maximal torus subgroup of the isometry of the sphere, $U(1)\times U(1) \subset SO(4)$, and the generator $R$ of the $U(1)_R$ R-symmetry.\footnote{There may also be a global flavor symmetry, in which case one would also introduce its Cartan subalgebra. We do not investigate this possibility here.}

The supercharges preserved by the background \eqref{eq:4d_SCI_Background} have anti-commutation relation
\begin{equation}
\label{eq:4d_BPS_Bound}
    \{ \cQ, \cQ^\dagger\} \= \EnergyOp - J_1 - J_2 - \frac{3}{2} R \, ,
\end{equation}
The spinors are anti-periodic around the $S^1$ if we require that the potentials satisfy
\begin{equation}
\label{eq:4d_SUSY_Potentials}
    \frac{\beta}{2\pi\ii} \bigl( 1 - 2 \Phi_R + \Omega_1 + \Omega_2 \bigr) \= 1+2n \, , \qquad n \in \Z \, .
\end{equation}
Because of this relation, we can manipulate the partition function on the background \eqref{eq:4d_SCI_Background} to take the form of a superconformal index:
\begin{equation}
\label{eq:4d_PF_SCI_Fili}
\begin{split}
    &\Tr_{\cH_{S^3}} \exp \Bigl( - \beta \EnergyOp + \beta \Omega_1 J_1 + \beta \Omega_2 J_2 + \beta \Phi_R R \Bigr) \\
    &\qquad \= \Tr_{\cH_{S^3}} \exp \Bigl( - \beta \{ \cQ, \cQ^\dagger \} + \beta \left( \Omega_1 - 1 \right) \left( J_1 + \tfrac{1}{2}R \right) + \beta \left(\Omega_2 - 1\right) \left(J_2 + \tfrac{1}{2} R \right) \\
    & \qquad \qquad \qquad \quad \quad \ \, + \frac{\beta}{2} \left( 2\Phi_R - 1 - \Omega_1 - \Omega_2 \right) R \Bigr)  \\
    &\qquad \= \cI_{\rm SC} \left( \frac{\beta}{2\pi\ii} \left( \Omega_1 - 1 \right), \frac{\beta}{2\pi\ii} \bigl(\, \Omega_2 - 1 \, \bigr) \right)
\end{split}
\end{equation}
where the superconformal index is 
\begin{equation}
\label{eq:4d_SCI_Definition}
\begin{split}
	\cI_{\rm SC}(\FirstParameterSCI, \SecondParameterSCI) \, \equiv \, \Tr_{\cH_{S^3}} (-1)^{ R} \e^{ - \beta \{ \cQ, \cQ^\dagger \} + 2\pi \ii \FirstParameterSCI \left( J_1 + \frac{1}{2}R \right) + 2\pi\ii \SecondParameterSCI \left( J_2 + \frac{1}{2} R \right) } \, ,
\end{split}    
\end{equation}
and we have identified
\begin{equation}
\label{eq:SUSY_SUGRA_ChemicalPotentials}
    \FirstParameterSCI \= \frac{\beta}{2\pi\ii} (\Omega_1 - 1) \,, \qquad \SecondParameterSCI \= \frac{\beta}{2\pi\ii} (\Omega_2 - 1) \, .
\end{equation}
Notice that we define the superconformal index as graded by the $R$-charge rather than by the angular momentum.
These two different gradings are related by a shift of the chemical potentials~$\FirstParameterSCI, \SecondParameterSCI$ by~$\pm 1$. 
The grading by the $R$-charge shows a growth of states consistent with the black hole in the ``Cardy-like'' limit~$\SecondParameterSCI\to 0$~\cite{Benini:2018ywd, Choi:2018hmj,Honda:2019cio,ArabiArdehali:2019tdm,Kim:2019yrz,Amariti:2019mgp,Cabo-Bizet:2019osg,GonzalezLezcano:2020yeb,Goldstein:2020yvj,Jejjala:2021hlt,Cassani:2021fyv,ArabiArdehali:2021nsx}.\footnote{Other saddles of the index of $\cN=4$ SYM~\cite{Cabo-Bizet:2019eaf,Cabo-Bizet:2020nkr} are dual to subleading gravity configurations that involve orbifolds of the internal $S^5$ \cite{Aharony:2021zkr}.}

It is clear that the background \eqref{eq:4d_SCI_Background} matches the conformal conditions in \eqref{eq:AdS5_BdryMetric}, and the relations \eqref{eq:4d_BPS_Bound} and \eqref{eq:4d_SUSY_Potentials} between conserved charges and thermodynamic potentials (respectively) match \eqref{eq:5d_AdSKN_BPSBound} and \eqref{eq:5d_AdSKN_ChemicalPotentials_Constraint}, so the supersymmetric supergravity solutions described earlier are good candidates to describe saddle points of the index \eqref{eq:4d_SCI_Definition} of the SCFT$_4$ just introduced. 
Indeed, the on-shell action of the supergravity solutions (obtained via background subtraction) is \cite{Cabo-Bizet:2018ehj}
\begin{equation}
    I \= \frac{4\pi^2 \ii}{27 G_5} \frac{\varphi_g^3}{\FirstParameterSCI_g \SecondParameterSCI_g} \, , 
\end{equation}
with $\varphi_g, \sigma_g$ and $\tau_g$ given below in \eqref{eq:5d_AdSKN_ReducedChemicalPotentials_taus}-\eqref{eq:5d_AdSKN_ReducedChemicalPotentials_varphi}. This matches the large-$N$ and ``Cardy-like'' limit of the superconformal index \eqref{eq:4d_SCI_Definition} of various families of four-dimensional superconformal field theories, with the appropriate identification of $N$ and $G_5$ (see \cite{Cassani:2025sim} for a review and a exhaustive list of references).

\subsection{Constraints on the parameters from the microscopic description}

We remark that the microscopic definition of the  supersymmetric index in~\eqref{eq:4d_SCI_Definition} imposes a constraint on the parameters. 
Namely, since the eigenvalues of $\{ \cQ, \cQ^\dagger \}$ are non-negative, we impose
\begin{equation}
\label{eq:5d_AdSKN_Convergence_Conditions_1}
    \Re \, \beta \, > \, 0
\end{equation}
for the trace on the full Hilbert space to be convergent. 
Further, since the eigenvalues of~$J_{1,2} + \frac{1}{2}R$ are non-negative on the BPS subspace, we impose
\begin{equation}
\label{eq:5d_AdSKN_Convergence_Conditions_2}
    \Im \FirstParameterSCI \,> \, 0 \, , \qquad \Im \SecondParameterSCI \, > \, 0 
\end{equation}
for the convergence of the 
trace~\cite{Aharony:2021zkr,Choi:2025lck}.

\smallskip

The a priori complex parameters $\beta$, $\FirstParameterSCI$ and $\SecondParameterSCI$ in \eqref{eq:SUSY_SUGRA_ChemicalPotentials} take specific values at the saddle points described by the gravitational solution, which we label by a ``$g$" subscript, and they are expressed in terms of the three real parameters $(a, r_+, r_\star)$: $\beta_g$ is found in \eqref{eq:SUSY_Thermodynamic_Potentials}, and the others read
\begin{equation}
\label{eq:5d_AdSKN_ReducedChemicalPotentials_taus}
\begin{split}
    \FirstParameterSCI_g 
    &\= \frac{\beta_g}{2\pi\ii} \bigl( \Omega_1 - \Omega_{1\star} \bigr) \\
    &\= \frac{ (1-a) ( r_+ \mp \ii b )}{ \ii \left(r_\star^2 - 3 r_+^2 \right) \mp 2 r_+ ( 1 + a + b ) } \,, \\[10pt]
    \SecondParameterSCI_g &\= \frac{\beta_g}{2\pi\ii} \bigl( \Omega_2 - \Omega_{2\star} \bigr) \\
    &\= \frac{ (1-b) ( r_+ \mp \ii a )}{ \ii \left(r_\star^2 - 3 r_+^2 \right) \mp 2 r_+ ( 1 + a + b ) } \,.
\end{split} 
\end{equation}
We can also introduce an additional reduced potential,
\begin{equation}
\label{eq:5d_AdSKN_ReducedChemicalPotentials_varphi}
    \varphi_g \, \equiv \, \frac{\beta_g}{2\pi\ii} \bigl( \Phi_e - \Phi_{e\star} \bigr)  
    \= \frac{1}{2} ( \pm 1 + \FirstParameterSCI_g + \SecondParameterSCI_g ) \,,
\end{equation}
where the second equality follows from~\eqref{eq:5d_AdSKN_ChemicalPotentials_Constraint}.

We can then represent the microscopic convergence conditions \eqref{eq:5d_AdSKN_Convergence_Conditions_1} and \eqref{eq:5d_AdSKN_Convergence_Conditions_2} on the space $(a, r_+, r_\star)$, together with the geometric constraints \eqref{eq:Geometric_Constraint_Degeneracy_v2} and \eqref{eq:Geometric_Constraint_rPlus_rStar}. We denote this region by
\begin{equation} 
\label{microregion}
	\textsf{micro} \= \{ (a,r_+,r_\star) \ \big| \ \text{Re} \, \beta_g > 0 \,, ~ \text{Im} \, \sigma_g > 0 \,, ~ \text{Im} \, \tau_g > 0 \} \,.
\end{equation}

Firstly, we note that, since~$r_+>r_\star$,  
the condition~\eqref{eq:5d_AdSKN_Convergence_Conditions_1}, with~$\beta_g$ given in~\eqref{eq:SUSY_Thermodynamic_Potentials}, is equivalent to~\eqref{eq:Geometric_Constraint_Degeneracy_v2}. 
Now we discuss the conditions~\eqref{eq:5d_AdSKN_Convergence_Conditions_2}. 
To this end, we introduce $x=X(r_\star)$ as the smallest real solution of the cubic equation
\begin{equation}
\label{eq:AdS5_CubicX}
    x^3 + x^2 \left(1-2 r_\star^2 \right) + x \left(1-2 r_\star^2 \right) - r_\star^4 -2 r_\star^2 \= 0 \, ,
\end{equation}
and~$y=Y(r_\star)$ as the smallest real solution of the cubic equation
\begin{equation}
\label{eq:AdS5_CubicY}
    y^3+y^2+y \left(2 r_\star^2+1\right)+r_\star^2 \= 0 \, .
\end{equation}
These equations arise from setting $\Im \FirstParameterSCI_g=0$ and~$\Im \SecondParameterSCI_g=0$, respectively, at $r_+=r_\star$. 
Looking at these conditions as equations defining $a(r_\star)$, one finds the solution $a=1$ (resp. $a = (r_\star^2-1)/2$, which is $b(a,r_\star)=1$), and the two solutions of 
equations~\eqref{eq:AdS5_CubicX} and~\eqref{eq:AdS5_CubicY}.

There are two ``interesting'' values for $r_\star$: $r_\star=\sqrt{3}$, which is the upper bound given in~\eqref{eq:Geometric_Constraint_Degeneracy_v2}, 
and~$X(r_\star)=1$ or,
equivalently,~$Y(r_\star)=\frac{1}{2}(r_\star^2-1)$, which is given by~$r_\star = \sqrt{2 \sqrt{3}-3} \equiv R_\star \approx 0.68$. 

For~$R_\star < r_\star < \sqrt{3}$, 
$\Im \FirstParameterSCI_g$ 
and $\Im \SecondParameterSCI_g$ are both positive in the domain in $(a, r_+, r_\star)$ defined by \eqref{eq:Geometric_Constraint_Degeneracy_v2} and \eqref{eq:Geometric_Constraint_rPlus_rStar}, i.e., the convergence of the microscopic trace is equivalent to the geometric conditions, as in the four-dimensional cases treated in \cite{BenettiGenolini:2025jwe}. 

For $0< r_\star < R_\star$ the conditions are more complicated, as we now discuss. Recall that we always impose~$r_+>r_\star$. 
Here, 
we find that $\Im \FirstParameterSCI_g$ 
and $\Im \SecondParameterSCI_g$ are both positive for the following three ranges of parameters, 
\begin{equation}
\label{eq:5d_AdSKN_RegionsPositivity}
\begin{split}
    &  \frac{r_\star^2 -1}{2} \, < \, a \, \leq \, Y(r_\star) \quad \text{and} \quad  r_+ \, > \, \sqrt{\frac{-6 a \left(a^2+a + 1 \right) + 3 (1-a) r_\star^2}{9(1+a)}} \, > \, r_\star \,,  \\
    &   Y(r_\star) \, < \, a \, \leq \, X(r_\star)  \,,   \\
    &   X(r_\star) \, < \, a \, < \, 1 \quad \text{and} \quad r_+ \, > \, \frac{\sqrt{6 a \left(a^2+a+ 1 \right)-6 r_\star^4-3 (1 - a)^2 r_\star^2}}{3(1+a)} \, > \, r_\star \, .
\end{split}
\end{equation}
Notice that the bounds on~$r_+$ by the expression in the square root in the first and third lines
cut out a region of the space allowed by the  geometric constraints.

\begin{figure}
    \centering
    \begin{overpic}[width=0.8\linewidth]{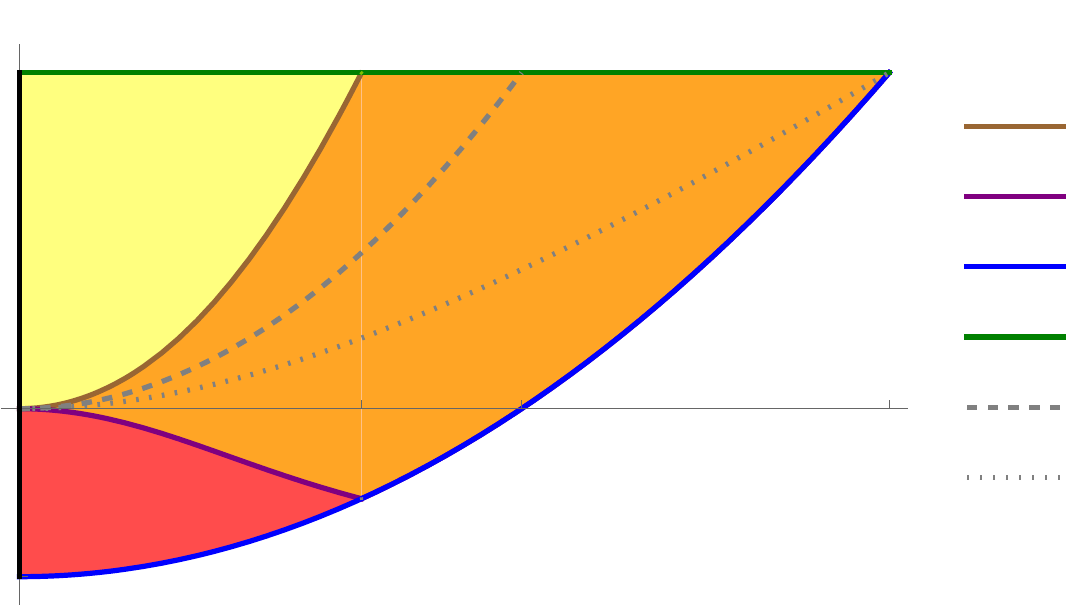}
        \put(4,182){\small $a$}
        \put(295,60){\small $r_\star$}
        \put(-2,168){\footnotesize $1$}
        \put(-4,60){\footnotesize $0$}
        \put(-16,8){\footnotesize $-0.5$}
        \put(170,56){\footnotesize $1$}
        \put(278,68){\footnotesize $\sqrt{3}$}
        \put(112,67){\footnotesize $R_\star$}
        \put(344,150){\footnotesize $X(r_\star)$}
        \put(344,128){\footnotesize $Y(r_\star)$}
        \put(344,107){\footnotesize $\frac{r_\star^2-1}{2}$}
        \put(344,84){\footnotesize $1$}
        \put(344,62){\footnotesize $r_\star^2$}
        \put(344,39){\footnotesize $\sqrt{1+r_\star^2}-1$}
        \put(150,173){\footnotesize $a=1$}
        \put(226,101){\footnotesize $b=1$}
        \put(140,128){\footnotesize $b=0$}
        \put(180,106){\footnotesize $b=a$}
    \end{overpic}
    \caption{\emph{Plot of geometric and  microscopic constraints~(cross-section at~$r_+=r_\star$)}. The figure shows the 
    the~$(r_\star,a)$-plane at the extremal point~$r_+=r_\star$.
    The region bounded by the horizontal green line~$a=1$, the blue line $a = (r_\star^2 -1)/2$ ($b=1$), and the vertical axis contains all the points allowed by the geometric constraints given in~\eqref{eq:Geometric_Constraint_Degeneracy_v2}. 
    The orange region is where~$\text{Im} \,\FirstParameterSCI_g >0$ and~$\Im \SecondParameterSCI_g >0$. The red region is where~$\text{Im} \,\FirstParameterSCI_g >0$ but~$\text{Im} \, \SecondParameterSCI_g <0$. The yellow region is where~$\text{Im} \, \SecondParameterSCI_g >0$ but~$\text{Im} \, \FirstParameterSCI_g <0$.
    The red-orange separator~$\Im \FirstParameterSCI_g = 0$ is given by~$a=X(r_\star)$. 
    The yellow-orange separator~$\Im  \SecondParameterSCI_g =0$ is given by~$a=Y(r_\star)$.  
    }
    \label{fig:5d_Imtau_BPS_v2}
\end{figure}

\begin{figure}
    \centering
    \scalebox{0.8}{\begin{overpic}[width=0.6\linewidth]{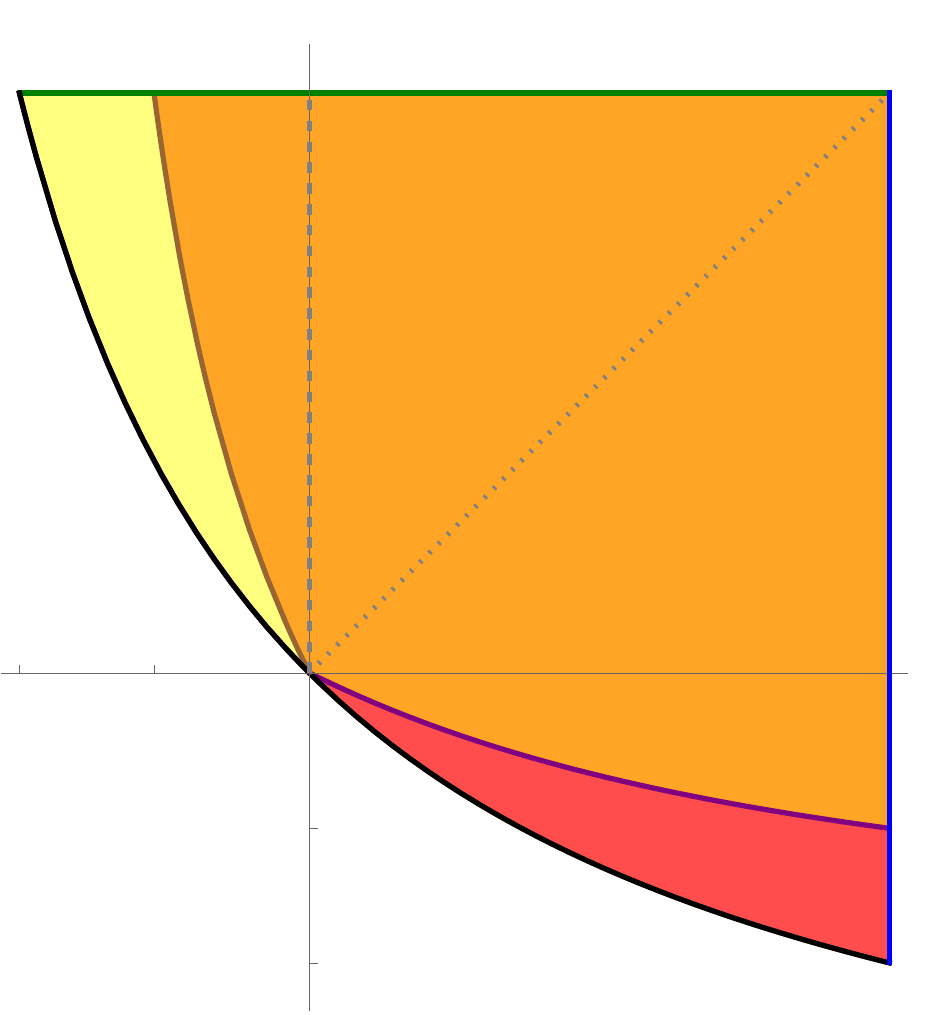}
        \put(83,270){\large $a$}
        \put(252,92){\large $b$}
        \put(-7,98){$-0.5$}
        \put(60,12){$-0.5$}
    \end{overpic}}
    \caption{\emph{Plot of geometric and  microscopic constraints~(cross-section at~$r_+=r_\star$)}. The figure shows the 
    the~$(b,a)$-plane at the extremal point~$r_+=r_\star$. 
    The boundaries of the region are at $r_\star = \sqrt{a+b+ab}=0$, $a=1$ and $b=1$.
    The color coding of the regions, and the colors and styles of the various curves are the same as in Figure \ref{fig:5d_Imtau_BPS_v2}.}
    \label{fig:5d_Imtau_BPS_ab}
\end{figure}

\begin{figure}
        \centering
        \begin{subfigure}[b]{0.475\textwidth}
            \centering
            \begin{overpic}[width=\textwidth]{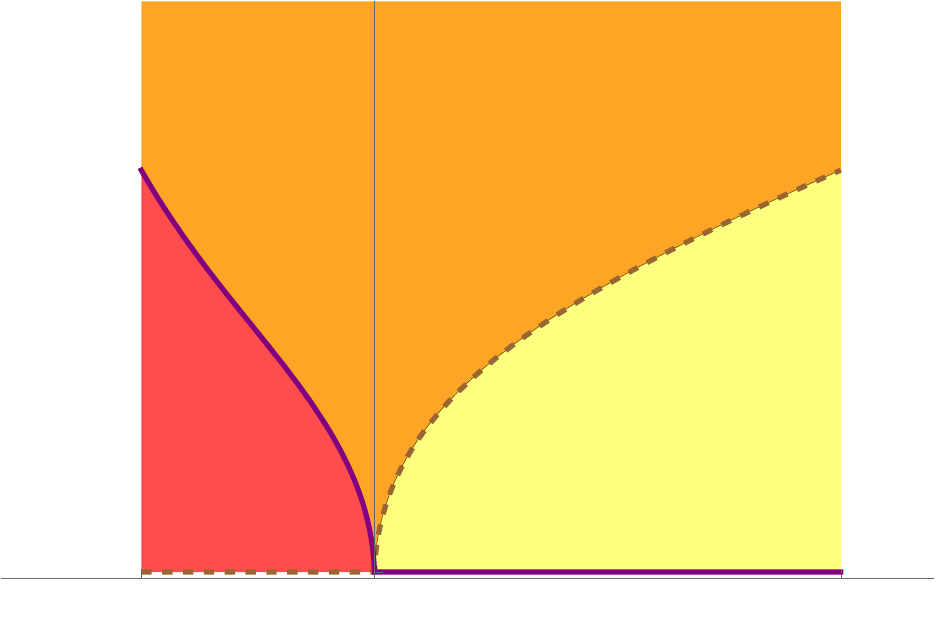}
                \put(79,137){\footnotesize $r_+$}
                \put(85,12){\footnotesize $r_\star$}
                \put(20,-3){\footnotesize $\frac{r_\star^2-1}{2}$}
                \put(60,0){\footnotesize $Y(r_\star)$}
                \put(83,0){\footnotesize $X(r_\star)$}
                \put(182,0){\footnotesize $1$}
                \put(206,7){\footnotesize $a$}
                \put(50,20){\footnotesize $\mc{U}_1$}
                \put(130,20){\footnotesize $\mc{U}_2$}
            \end{overpic}
            \caption{{\small $r_\star = 0.01$}}    
            \label{fig:CrossSections_001}
        \end{subfigure}
        \hfill
        \begin{subfigure}[b]{0.475\textwidth}  
            \centering 
            \begin{overpic}[width=\textwidth]{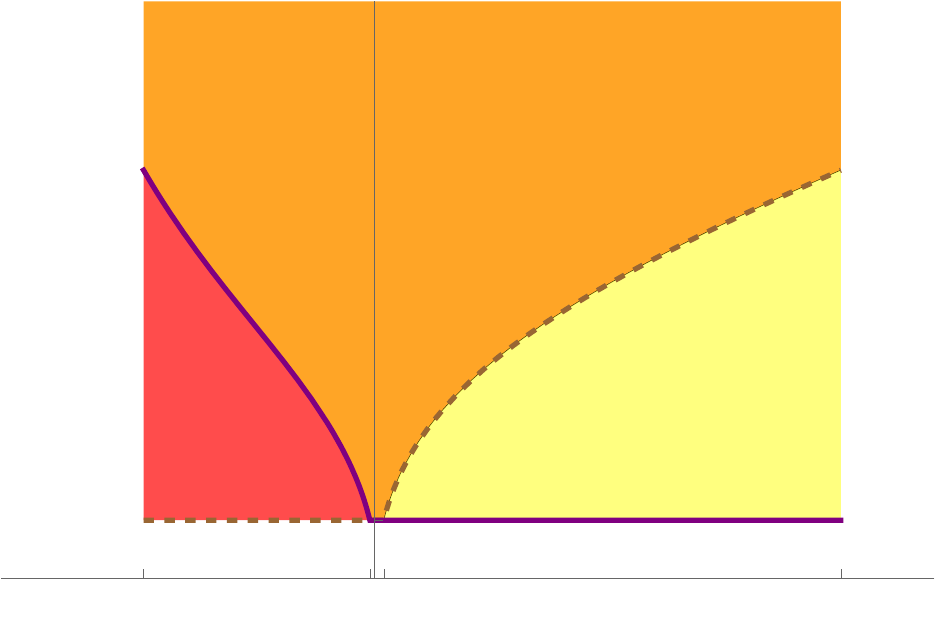}
                \put(79,137){\footnotesize $r_+$}
                \put(85,14){\footnotesize $r_\star$}
                \put(21,-3){\footnotesize $\frac{r_\star^2-1}{2}$}
                \put(60,0){\footnotesize $Y(r_\star)$}
                \put(84,0){\footnotesize $X(r_\star)$}
                \put(182,0){\footnotesize $1$}
                \put(206,7){\footnotesize $a$}
                \put(50,30){\footnotesize $\mc{U}_1$}
                \put(130,30){\footnotesize $\mc{U}_2$}
            \end{overpic}
            \caption{{\small $r_\star = 0.1$}}     
            \label{fig:CrossSections_01}
        \end{subfigure}
        \vskip\baselineskip
        \begin{subfigure}[b]{0.475\textwidth}   
            \centering 
            \begin{overpic}[width=\textwidth]{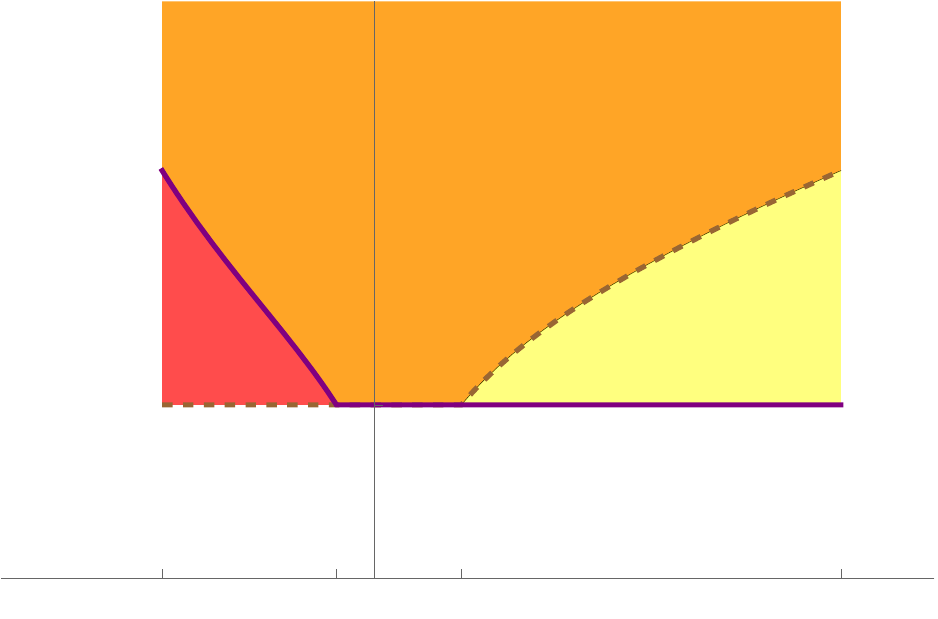}
                \put(79,137){\footnotesize $r_+$}
                \put(85,40){\footnotesize $r_\star$}
                \put(24,-3){\footnotesize $\frac{r_\star^2-1}{2}$}
                \put(60,0){\footnotesize $Y(r_\star)$}
                \put(92,0){\footnotesize $X(r_\star)$}
                \put(182,0){\footnotesize $1$}
                \put(206,7){\footnotesize $a$}
                \put(46,55){\footnotesize $\mc{U}_1$}
                \put(140,55){\footnotesize $\mc{U}_2$}
            \end{overpic}
            \caption{{\small $r_\star = 0.3$}}     
            \label{fig:CrossSections_03}
        \end{subfigure}
        \hfill
        \begin{subfigure}[b]{0.475\textwidth}   
            \centering 
            \begin{overpic}[width=\textwidth]{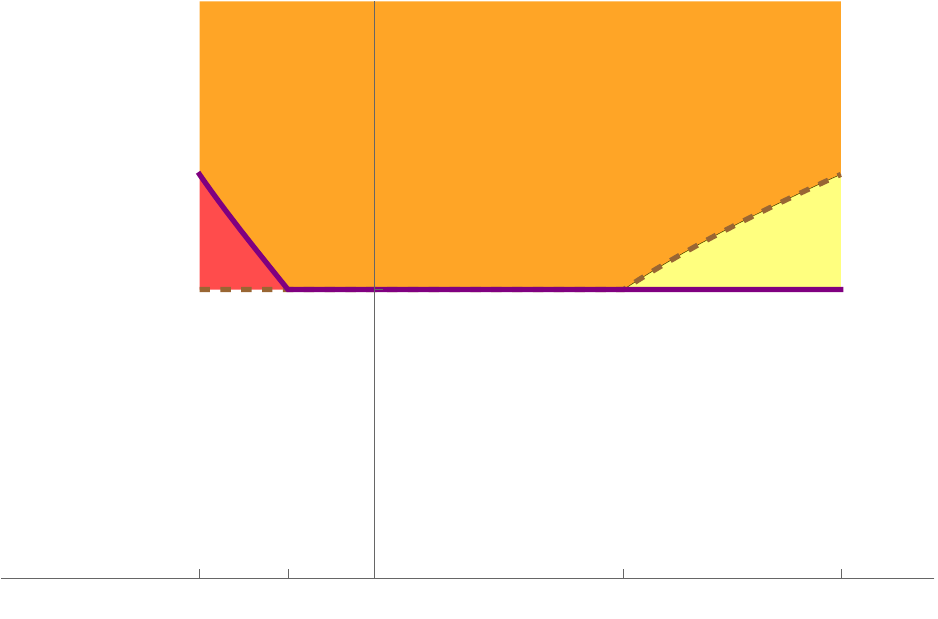}
                \put(79,137){\footnotesize $r_+$}
                \put(85,64){\footnotesize $r_\star$}
                \put(34,-3){\footnotesize $\frac{r_\star^2-1}{2}$}
                \put(56,0){\footnotesize $Y(r_\star)$}
                \put(128,0){\footnotesize $X(r_\star)$}
                \put(182,0){\footnotesize $1$}
                \put(206,7){\footnotesize $a$}
                \put(46,75){\footnotesize $\mc{U}_1$}
                \put(162,75){\footnotesize $\mc{U}_2$}
            \end{overpic}
            \caption{{\small $r_\star = 0.5$}}       
            \label{fig:CrossSections_05}
        \end{subfigure}
        \caption{\emph{Plot of geometric and  microscopic constraints (cross-section at fixed~$r_\star$)}. The figure shows the 
        the~$(a,r_+)$-plane at fixed $r_\star = 0.01, 0.1, 0.3, 0.5$.
        The colored region (all colors) represents the region allowed by geometric constraints given 
        in~\eqref{eq:Geometric_Constraint_Degeneracy_v2}. 
        The microscopic constraints described in~\eqref{eq:5d_AdSKN_RegionsPositivity} divide the region into three parts, as in Figure~\ref{fig:5d_Imtau_BPS_v2} with the same color coding: red = $\text{Im} \, \FirstParameterSCI_g >0$ and $\text{Im} \, \SecondParameterSCI_g < 0$, yellow = $\text{Im} \, \SecondParameterSCI_g >0$ and $\text{Im} \, \FirstParameterSCI_g < 0$,
        orange = both~$\text{Im} \, \FirstParameterSCI_g , \, \text{Im} \, \SecondParameterSCI_g >0$.}
        \label{fig:CrossSections}
\end{figure}
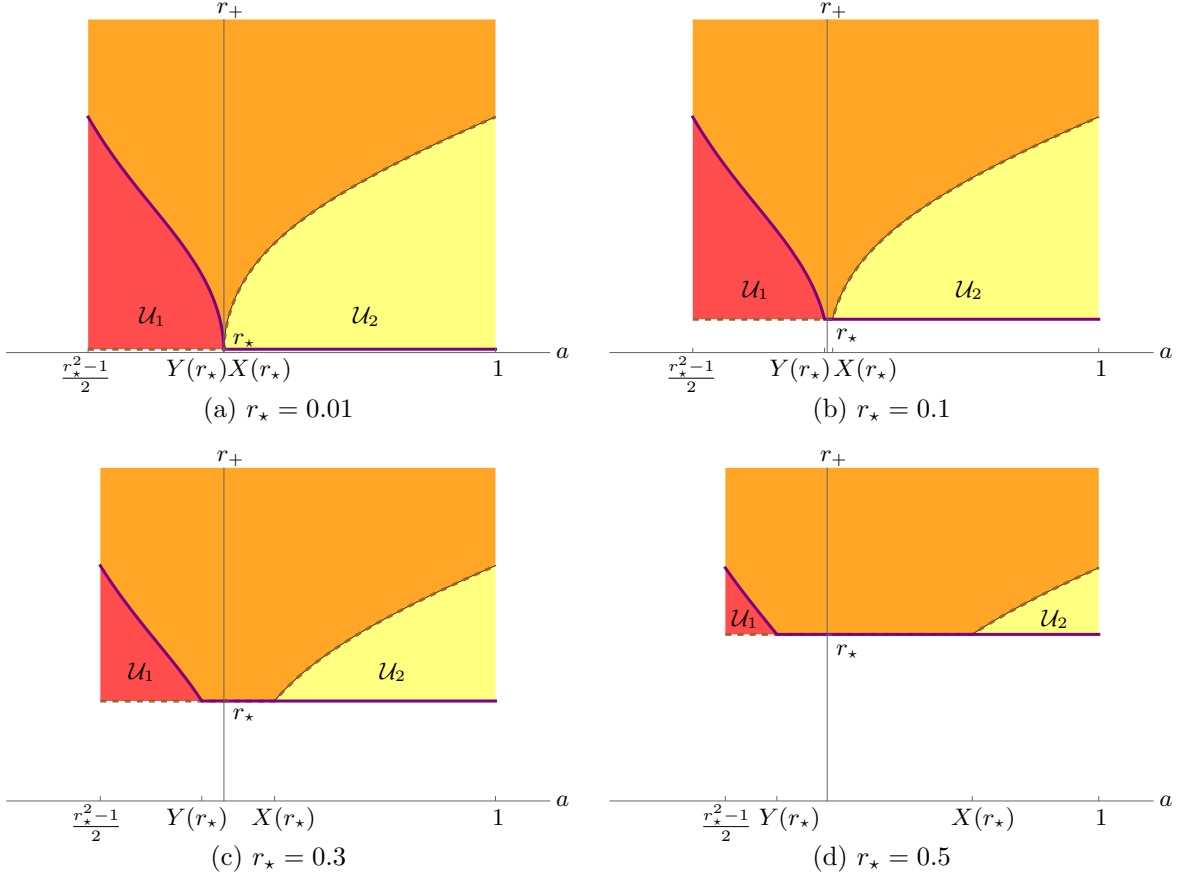

In Figure~\ref{fig:5d_Imtau_BPS_v2} we represent the projection of these regions on the $(a,r_\star)$-plane, and in Figure~\ref{fig:CrossSections} we represent them on a section of the $(a,r_+)$ plane at fixed $r_\star$. 
For completeness, we also represent the  
content of Figure~\ref{fig:5d_Imtau_BPS_v2} in Figure~\ref{fig:5d_Imtau_BPS_ab}. 
The latter figure is in the $(b,a)$ plane, in which the symmetry in $a\leftrightarrow b$ is manifest. 
In both these figures, the region with any color represents points where the geometric constraints are satisfied. 
The region in orange is the one where the microscopic constraints are satisfied. 
The complement of the orange region is the region where the geometric constraints are satisfied, but the microscopic ones are not, and is given by 
\begin{equation}
\label{eq:5d_AdSKN_RegionsKSW_NotPositivity}
\begin{split}
    \mc{U} &\= \mc{U}_1 \, \cup \,  \mc{U}_2 \,, \\
    \mc{U}_1 & \=  \Bigg\{ \frac{r_\star^2 -1}{2} \, < \, a \, \leq \, Y(r_\star) \ \text{and} \ 
    \sqrt{\frac{-6 a \left(a^2+a+1\right) + 3 (1-a) r_\star^2}{9(1+a)}} \, > \, r_+ \, > \, r_\star \Bigg\} \,, \\
    \mc{U}_2 & \= \Bigg\{ X(r_\star) \, < \, a \, < \, 1 \ \text{and} \ 
    \frac{\sqrt{6 a \left(a^2+a + 1\right)-6 r_\star^4-3 (1-a)^2 r_\star^2}}{3(1+a)} \, > \, r_+ \, > \, r_\star \Bigg\} \, .
\end{split} 
\end{equation}
As evident from Figure~\ref{fig:5d_Imtau_BPS_ab}, $\cU$ is entirely contained in the region 
where $ab<0$ (within the region allowed by the geometric constraints).

\smallskip

It is interesting to see what happens as we take the angular momenta to be equal: setting $a=b$ implies that $r_\star^2 = a(a + 2)$, which is only consistent with  
our assumptions if $a>0$, so that the family of solutions with equal angular momenta is parametrized by $(a,r_+)$ with
\begin{equation}
\label{eq:5d_AdS_KN_Constraints_EqualVelocity}
    0 \, < \, a \, < \, 1 \, , \qquad r_+ \, > \, \sqrt{a(a+2)} \, .
\end{equation}
We find that when these two conditions hold, then $\abs{\Omega_1}=\abs{\Omega_2} < 1$. 
The inverse temperature and the only independent chemical potential can be found from  \eqref{eq:SUSY_Thermodynamic_Potentials} and \eqref{eq:5d_AdSKN_ReducedChemicalPotentials_taus}, respectively: 
\begin{equation}
\label{eq:AdS5_Potentials_EqualVelocity}
\begin{split}
    \beta_g &\= 2 \pi \frac{  r_+ \left( 3 r_+^2 + a (a + 2) (2 a - 1 ) \right) \pm \ii  \left(a^2 (a+2 )^2+r_+^2 \left(3 a^2+4 a +2\right)\right) }{\left((a+2 )^2+9 r_+^2\right) \left(r_+^2-a (a+2 )\right)} \, , \\
    \FirstParameterSCI_g &\= \frac{ (1-a) ( \mp ( 2+a ) + 3 \ii r_+ ) }{ (2+a)^2 + 9 r_+^2} \, .
\end{split}
\end{equation}
In contrast to the case of unequal angular momenta, we see that the conditions $\Re\beta_g >0$ and \hbox{$\Im\FirstParameterSCI_g>0$} \textit{both} hold in the region \eqref{eq:5d_AdS_KN_Constraints_EqualVelocity}. 
Therefore, the geometric constraints and the microscopic constraints are equivalent. 
This is also clear in Figures~\ref{fig:5d_Imtau_BPS_v2} and~\ref{fig:5d_Imtau_BPS_ab}, 
where one sees that the curve $b=a$ (in gray dotted) lies entirely in the orange region.

\section{The KSW criterion and the \texorpdfstring{AdS$_5$}{AdS5} black hole index \label{sec:KSWAdS5}}

In the previous section, we introduced the complex supersymmetric supergravity solutions that are relevant for the saddle point evaluation of the gravitational description of a four-dimensional superconformal index. They depend on three parameters $(a, r_+, r_\star)$, and we argued that while a good definition of the geometry restricts their ranges, convergence of the microscopic description of the superconformal index restricts them even further. 
In this section, we show that the restrictions on the parameters of the gravity solution imposed by 
the KSW criterion are equivalent to those imposed by the convergence of the microscopic description of the index.

\subsection{KSW criterion}

Consider a smooth manifold $M$ with a complex-valued, non-degenerate symmetric tensor $g$. 
Kontsevich and Segal~\cite{Kontsevich:2021dmb} argue that it is consistent to define a quantum field theory on $(M,g)$ provided at each point $p\in M$, 
the complex-valued quadratic form induced by the metric on $\Lambda^qT_p^*M$ has nowhere $\det g \in \R^-$, 
in which case we choose the branch of the square root such that $\text{Re} \sqrt{\det g} > 0$ everywhere. With this choice, the metric satisfies 
\begin{equation}
\label{eq:Allowability_RealPart}
    \Re \left( \sqrt{\det g} \, g^{i_1 j_1} g^{i_2 j_2} \cdots g^{i_q j_q} F_{i_1 i_2\cdots i_q} F_{j_1 j_2\cdots j_q} \right) \, > \, 0 \qquad 0 \leq q \leq D
\end{equation}
for any real, non-zero $q$-form $F$. This guarantees that the real part of the quadratic form, or, more physically, the real part of the kinetic term, is positive-definite. 
They then show that requiring this positive-definiteness is equivalent to saying that there is a basis of the real
vector space $T_pM$ in which the quadratic form on $T_pM$ induced by $g_p$ can be written as
\begin{equation}
\label{eq:QuadraticForm_Diagonal}
    g_p \= \sum_{i=1}^D \lambda_i \, y_i^2 \,,
\end{equation}
where the $y_i$ are coordinates 
in that basis, and the $\lambda_i$ are non-zero complex numbers, not on the negative real axis, such that 
\begin{equation}
\label{eq:Allowability_Eigenvalues}
    \sum_{i=1}^D \abs{\Arg\lambda_i } \, < \, \pi \, ,
\end{equation}
where $\Arg \in (-\pi,\pi]$ is the principal value of the argument. Witten suggested in \cite{Witten:2021nzp} that a gravity solution is a good saddle point of the gravitational path integral provided it is \textit{allowable} according to the Kontsevich--Segal criterion \eqref{eq:Allowability_RealPart}, so we refer 
to this condition as the KSW criterion. This builds on earlier work in quantum cosmology~\cite{Halliwell:1989dy,Louko:1995jw}.

\smallskip

To verify the restrictions imposed by the KSW criterion on the five-dimensional complex solutions introduced in the previous section, 
we impose the condition~\eqref{eq:Allowability_RealPart} everywhere on the underlying manifold $\R^2 \times S^3$. 
We denote the set of 3-parameter metrics for which this holds by
\begin{equation}
	\textsf{KSW} \, \equiv \, \{ (a,r_+,r_\star) \ \big| \ \eqref{eq:Allowability_RealPart} \text{ holds everywhere on } M \} \, .	
\end{equation}
We were not able to analytically characterize the set \textsf{KSW}, and hence we proceed as follows. 
Firstly, we explore analytically the constraints of the KSW criterion 
on the metric at the conformal boundary, 
by considering the first non-trivial order in the expansion near the boundary. 
We then move to a numerical evaluation of the KSW criterion in the bulk. 
In order to do so, we  re-express the KSW criterion in a form that we codify as the following simple
algorithm, which determines whether a given complex-valued metric $g$ on $M$ is KSW-allowable. \\ 
Regard the components $g_{\mu \nu}$ of~$g$ in some real coordinate basis as a $D \times D$ matrix.
Then, $g$ is KSW-allowable if and only if for 
each $p \in M$ the metric~$g_p$ obeys the following three conditions:
\begin{enumerate}
	\item $\det g_p \notin \mathbb{R}^-$, in which case define $\text{Re} \sqrt{\det g_p} \, > 0$,
	\item the matrix $A_p \equiv \Re \left( \sqrt{\det g_p} \, g^{-1}_p \right)$ is positive definite,
	\item the following condition holds
\begin{equation} \label{KSWeigenvalueEq}
	\text{Tr} \, \bigl| \, \text{arg} \bigl( g_p A_p \bigr)  \bigr| \; \equiv \;  
    \sum_{i=1}^D \, \bigl| \,\text{arg} \bigl( \mu_i ( g_p A_p ) \bigr) \bigr| \, < \, \pi \,,
\end{equation}
where $\mu_i \left( g_p A_p \right)$, $i = 1, \dots, D$, are the eigenvalues of $g_pA_p$.
\end{enumerate}
We prove this slight re-characterization of~\cite{Kontsevich:2021dmb,Witten:2021nzp} in Appendix~\ref{appA}. 
Note the difference between~\eqref{eq:Allowability_Eigenvalues} and~\eqref{KSWeigenvalueEq}: 
the latter condition is imposed on the  \textit{eigenvalues} $\mu_i$ of the \textit{matrix}~$g_p A_p$, whereas the $\lambda_i$ appearing in the formula \eqref{eq:Allowability_Eigenvalues} are not eigenvalues, as $g_p$ there is diagonalized by congruence, as appropriate for quadratic forms.
The above algorithm is practical because it avoids having to change basis to put $g_p$ in the diagonal form \eqref{eq:QuadraticForm_Diagonal} by congruence. 
It can be used to numerically rule out complex-valued metrics according to KSW (or to gain confidence that they satisfy KSW, as we do later on), or, if one is powerful enough, to prove analytically that a given metric satisfies KSW. Notice that 
\begin{itemize}
    \item if $g_{\mu \nu}$ is diagonal and the conditions 1~and 2~above are satisfied, then in~\eqref{KSWeigenvalueEq} one can replace $g_p A_p$ by $g_p$.
    To see this, note that in this case, $A_p$ is a positive definite diagonal matrix, and so  the eigenvalues of $g_p A_p$ are related to those of $g_p$ by rescaling by a positive real number. 
    Hence they have the same phase, which implies that~$\text{Tr} \, | \text{arg} ( g_p A_p )  | = \text{Tr} \, |  \text{arg} ( g_p )  |$.
    \item the algorithm is independent of the real coordinate basis in which we choose to express $g_{\mu \nu}$ as a matrix. 
    To see this, note that the metric in a new basis would take the form~$\tilde{g} = M^\top g M$ for a real invertible matrix $M$. 
    Therefore $\det \tilde{g} = (\det M)^2 \det g$, from which it follows that the Condition~1 is satisfied for $g$ 
    if and only if it is satisfied for $\tilde{g}$. 
    Similarly for Condition~2, we have $A = |\det M|^{-1} \, M \tilde{A} M^\top$ so that $A$ is positive definite if and only if $\tilde{A}$ is. 
    Finally, $\tilde{g} \tilde{A} = |\det M| \, M^\top g A \, (M^\top)^{-1}$, so that the eigenvalues of $\tilde{g} \tilde{A}$ are equal to the eigenvalues of $g A$ up to rescaling by a positive number. This proves the equivalence of Condition~3 for both ways of expressing the metric.
\end{itemize}
Concretely, for our purposes, this algorithm will allow us to provide numerical evidence that the constraints imposed by the KSW criterion are strongest at the conformal boundary (where we have analytical control) and become weaker as one moves into the bulk. We explain and discuss this in more detail at the end of this section.

\subsection{KSW criterion at the conformal boundary}
\label{sec:bulkbdry}

The metric near the conformal boundary at leading order in the Fefferman--Graham coordinate~$z$ takes the form~\eqref{eq:AdS5_BdryMetric}. 
Given that~$z, \vartheta \in\R$, we can, for the purpose of verifying the KSW criterion in this region, focus on the following complex line element on~$\tilde{M} = T^3$:
\begin{equation}
\label{eq:Bdry_LineElement}
\begin{split}
    \rd s^2_3 &\= \beta^2 ( 1 - \Omega_1^2 \sin^2\BdryTheta - \Omega_2^2 \cos^2\BdryTheta ) \rd \Periodictau^2 + \sin^2\BdryTheta \, \rd\phiCanonicalPeriodicities^2 + \cos^2\BdryTheta \, \rd\psiCanonicalPeriodicities^2 \\
    & \qquad - 2 \ii \beta \rd \Periodictau ( \Omega_1 \sin^2\BdryTheta \, \rd\phiCanonicalPeriodicities + \Omega_2 \cos^2\BdryTheta \, \rd\psiCanonicalPeriodicities) \, ,
\end{split}
\end{equation}
where the complex-valued $\beta$, $\Omega_1$ and $\Omega_2$ are given in~\eqref{eq:SUSY_Thermodynamic_Potentials}. In the 
real basis $\{ \rd\Periodictau, \, \rd\phiCanonicalPeriodicities, \, \rd\psiCanonicalPeriodicities \}$, the metric is represented by the matrix  
\begin{equation}
\label{eq:gTilde0}
    \tilde{g} \, \equiv \, \begin{pmatrix}
        \beta^2 ( 1 - \Omega_1^2 \sin^2\BdryTheta - \Omega_2^2 \cos^2\BdryTheta ) & - \ii \beta \Omega_1 \sin^2\BdryTheta & - \ii \beta \Omega_2 \cos^2\BdryTheta \\
        - \ii \beta \Omega_1 \sin^2\BdryTheta & \sin^2\BdryTheta & 0 \\
        - \ii \beta \Omega_2 \cos^2\BdryTheta & 0 & \cos^2\BdryTheta
    \end{pmatrix} \, .
\end{equation}
We write the boundary KSW criterion as 
\begin{equation}
	\textsf{BKSW} \, \equiv \, \{ (a,r_+,r_\star) \ \big| \ \eqref{eq:Allowability_RealPart} \text{ for } \tilde{g} \text{ holds everywhere on } \tilde{M} \} \, .
\end{equation}
It is clear by construction that 
\begin{equation} 
\label{trivialinclusion}
	\textsf{KSW} \subseteq \textsf{BKSW} \,,
\end{equation}
because $\textsf{KSW}$ imposes more conditions on $(a,r_+,r_\star)$ than $\textsf{BKSW}$.

\smallskip

The condition \eqref{eq:Allowability_RealPart} with $q=0$ is just $\Re \sqrt{\det \tilde{g}} = \sin \BdryTheta \cos\BdryTheta \, \Re \beta > 0$.
By assumption,~$\BdryTheta \in [0,\pi/2]$, and so the KSW condition with $q=0$ reduces to~$\Re\beta > 0$. 
Notice that this is the same as the microscopic convergence condition~\eqref{eq:5d_AdSKN_Convergence_Conditions_1}, 
which, as already pointed out, 
is equivalent to the constraints~\eqref{eq:Geometric_Constraint_Degeneracy_v2} on the parameters of the gravity solution.

\smallskip

The KSW criterion with $q=1$ requires checking the positive-definiteness of $\Re W \equiv \Re \sqrt{\det \tilde{g}} \, \tilde{g}^{-1}$
\begin{equation}
\label{eq:ReW0}
    \Re W \= \cos\BdryTheta \sin\BdryTheta \begin{pmatrix}
        \Re 1/\beta & - \Im \Omega_1 & -\Im \Omega_2 \\
        - \Im \Omega_1 & \Re\left[ \beta( \frac{1}{\sin^2\BdryTheta} - \Omega_1^2) \right] & - \Re (\beta \Omega_1\Omega_2) \\
        - \Im \Omega_2 & - \Re( \beta\Omega_1 \Omega_2 ) & \Re \left[ \beta( \frac{1}{\cos^2\BdryTheta} - \Omega_2^2) \right]
    \end{pmatrix}  \, .
\end{equation}
This can be done using Sylvester's criterion, that is, by checking that the leading principal minors of $\Re W$ are positive. 
The criteria must hold for all~$\vartheta \in [0,\pi/2]$. 
The first principal minor is positive because $\Re\beta >0$ from the KSW criterion for $q=0$, which is equivalent to the condition~$\Re \beta^{-1} > 0$.
Positivity of the second principal minor, written in terms of~$\FirstParameterSCI$, 
is equivalent to the inequality 
\begin{equation}
	\left( \Re \beta \right)^2 - \sin^2 \BdryTheta \left( \Re \beta - 2 \pi \, \Im \FirstParameterSCI \right)^2 \, > \, 0 \, ,
\end{equation}
which is strongest at $\BdryTheta = \pi/2$, where it is equivalent to
\begin{equation}
    \Im\FirstParameterSCI \left( \frac{1}{\pi}\Re\beta - \Im \FirstParameterSCI \right) \, > \, 0 \, .
\end{equation}
Since $\Re\beta>0$, this holds if and only if
\begin{equation}
\label{eq:bKSW_sigma}
    \frac{1}{\pi}\Re \beta \, > \, \Im\FirstParameterSCI \, > \, 0 \, .
\end{equation}
At first sight, this inequality seems too strong, as it seems to rule out a region of the parameter space that is 
allowed by the microscopic constraints~\eqref{eq:5d_AdSKN_Convergence_Conditions_1} 
and~\eqref{eq:5d_AdSKN_Convergence_Conditions_2}. 
However, one should recall that for the complex gravitational solution we are considering, 
$\beta$,~$\FirstParameterSCI$ and~$\SecondParameterSCI$ are not
independent complex numbers. 
Indeed, once we take into account the particular values~$\beta_g$,~$\FirstParameterSCI_g$ and~$\SecondParameterSCI_g$ given in~\eqref{eq:SUSY_Thermodynamic_Potentials} and~\eqref{eq:5d_AdSKN_ReducedChemicalPotentials_taus}, one can check that
\begin{equation} \label{eq:betasigineq1}
    \frac{1}{\pi}\Re \beta_g \, < \, \Im\FirstParameterSCI_g \quad \Rightarrow \quad \Im \SecondParameterSCI_g < 0 \,.
\end{equation}
We return to this thought presently. 

Now we move to the positivity of~$\det \Re W$ itself, which expressed in terms of~$\FirstParameterSCI$ and~$\SecondParameterSCI$ leads to the inequality
\begin{equation}
\Re\beta \times \bigl(\, \sin^2\BdryTheta  \Im \FirstParameterSCI  (\Re\beta -\pi  \Im\FirstParameterSCI )+ \cos^2\BdryTheta  \Im\SecondParameterSCI (\Re\beta -\pi  \Im\SecondParameterSCI ) \, \bigr) \, > \, 0 \, .
\end{equation}
This expression takes the form~$f = b + (a-b) x$ 
for certain $a$, $b$ and $x=\sin^2\BdryTheta$. 
The positivity of this linear function is guaranteed if~$f$ is positive at the edges of the interval $x=\sin^2\BdryTheta \in [0,1]$ that is, if
\begin{equation}
    \Im\FirstParameterSCI \, \Bigl(\, \frac{1}{\pi}\Re\beta - \Im \FirstParameterSCI \Bigr) \, > \, 0 \, , 
    \qquad \Im\SecondParameterSCI \, \Bigl( \,\frac{1}{\pi} \Re\beta - \Im \SecondParameterSCI \Bigr) \, > \, 0 \, .
\end{equation}
The first condition is guaranteed by \eqref{eq:bKSW_sigma}. For the second condition, the only possibility consistent with $\Re\beta >0$ is
\begin{equation}
\label{eq:bKSW_tau}
    \frac{1}{\pi}\Re \beta \, > \, \Im\SecondParameterSCI \, > \, 0 \, .
\end{equation}
As above, one can check that for the particular values of~$\beta_g$,~$\FirstParameterSCI_g$~~$\SecondParameterSCI_g$ given in~\eqref{eq:SUSY_Thermodynamic_Potentials} 
and~\eqref{eq:5d_AdSKN_ReducedChemicalPotentials_taus}
\begin{equation} \label{eq:betatauineq1}
    \frac{1}{\pi}\Re \beta_g \, < \, \Im\SecondParameterSCI_g \quad \Rightarrow \quad \Im \FirstParameterSCI_g < 0 \, .
\end{equation}

Now we put together the above two thoughts leading to~\eqref{eq:betasigineq1} and~\eqref{eq:betatauineq1}, respectively. 
We see that the regions of parameter space that would be removed by the first inequalities in~\eqref{eq:bKSW_sigma} and~\eqref{eq:bKSW_tau} 
are removed by requiring that~$\Im \FirstParameterSCI_g > 0 $ \textit{and}~$\Im \SecondParameterSCI_g > 0 $. 
We conclude that the KSW criterion applied to the leading term of the near-boundary expansion of 
the metric of the complex gravitational saddle is equivalent to the microscopic criterion, i.e,
\begin{equation} 
\label{nontrivialresult1}
	\textsf{BKSW} \= \textsf{micro} \,.
\end{equation}
By \eqref{trivialinclusion}, this already shows that the full KSW criterion is at least as strong as the microscopic convergence criteria,
\begin{equation} 
\label{trivialimplication1}
	\textsf{KSW} \subseteq \textsf{micro} \,.
\end{equation}

\subsection{KS criterion and index of the dual \texorpdfstring{$4d$}{4d} supersymmetric QFT}


Before moving to the next step in our application of the KSW criterion to the gravitational saddle, 
we take an instructive detour.
Recall that the criterion~\eqref{eq:Allowability_RealPart} was originally proposed by Kontsevich and Segal~\cite{Kontsevich:2021dmb} 
as a consistency criterion for a quantum field theory on a complex metric background 
(rather than quantum gravity). 
From this point of view, it is natural to enquire about the result of applying the Kontsevich--Segal (KS) criterion directly to 
the gauge theory holographically dual to the AdS$_5$ gravity that we have discussed so far. 

In the spirit of this paper, we 
specialize to the supersymmetric index of a superconformal field theory living on the 4-dimensional
background~\eqref{eq:4d_SCI_Background}. 
In this context, and without specifying any more details of the boundary theory, we ask: 
what are the conditions imposed 
on $\beta$, $\FirstParameterSCI$ and $\SecondParameterSCI$ 
by~\eqref{eq:Allowability_RealPart} applied to the 4-dimensional theory? 

Since we are assuming that the coordinates are real, for the purpose of checking allowability, we can focus on 
the line element~\eqref{eq:Bdry_LineElement} rather than considering the four-dimensional~\eqref{eq:4d_SCI_Background}, 
and therefore the analysis and conclusions of the previous section continue to be valid.
In other words, the criterion~\eqref{eq:Allowability_RealPart}, 
interpreted as a condition for the consistency of a quantum field theory, imposes the inequalities~\eqref{eq:bKSW_sigma} and~\eqref{eq:bKSW_tau} on the parameters, which we repeat here: 
\begin{equation}
    \label{eq:bdryts}
    \frac{1}{\pi}\Re \beta \, > \, \Im\FirstParameterSCI \, > \, 0 \,, \qquad 
    \frac{1}{\pi}\Re \beta \, > \, \Im\SecondParameterSCI \, > \, 0 \, .
\end{equation}

We are then led to the same small puzzle already raised in the gravitational context: the inequalities above are stronger than the microscopic constraints~\eqref{eq:5d_AdSKN_Convergence_Conditions_1} 
and~\eqref{eq:5d_AdSKN_Convergence_Conditions_2}, so how do we recover the latter and get rid of the first inequalities in~\eqref{eq:bdryts} (i.e.~$\frac{1}{\pi}\Re \beta > \Im\FirstParameterSCI, \Im\SecondParameterSCI$)? In the analysis of the previous section, we remarked that the $5d$ bulk gravitational saddle has additional geometric properties which relate the values~$\beta_g$ to~$\FirstParameterSCI_g$, $\SecondParameterSCI_g$. 
These made the first inequalities in~\eqref{eq:bdryts} redundant, and gave the result~\eqref{nontrivialresult1}, namely $\textsf{BKSW} = \textsf{micro}$. 
However, the background in~\eqref{eq:4d_SCI_Background} \textit{per se} does not  have these properties, 
and the three complex parameters $\beta, \sigma, \tau$ should be taken to be independent. 
In this case, to resolve the puzzle we should recall that the KS criterion pertains to the convergence of the thermal partition function on~$S^3$, whereas the microscopic constraints arise from the convergence of the trace defining the superconformal index, which involves a sum over a proper subset of the Hilbert space, and so leads to a weaker condition than~\eqref{eq:bdryts}. 

\smallskip

To be more concrete, we  re-express~\eqref{eq:bdryts} in terms of~$\Omega_1$ and~$\Omega_2$ using~\eqref{eq:SUSY_SUGRA_ChemicalPotentials} as
\begin{equation}
\label{eq:KS_S3_PartitionFunction}
    \Re\beta \, > \, 0 \, , \qquad \abs{\frac{\Re(\beta\Omega_1)}{\Re\beta}} \, < \, 1 \, , \qquad \abs{\frac{\Re(\beta\Omega_2)}{\Re\beta}} \, < \, 1 \, .
\end{equation} 
As already pointed out, these conditions are the result of the application of the KS criterion to the background~\eqref{eq:4d_SCI_Background}, 
so they guarantee the positive-definiteness of the kinetic operator.
Equivalently, they guarantee the convergence of the following partition function for any CFT on $S^3$
\begin{equation}
\label{eq:KS_ThermalPF}
    Z(\beta, \Omega_1, \Omega_2) \= \Tr_{\cH_{S^3}}\exp \left( - \beta E + \beta \Omega_1 J_1 + \beta \Omega_2 J_2 \right) \, .
\end{equation}
As a matter of fact, the same conditions are those found by applying the KSW criterion to a non-supersymmetric asymptotically AdS$_5$ black hole solution. 
This bulk/boundary consistency is the extension to complex~$\beta$ of the remarks made by Witten in~\cite{Witten:2021nzp}, 
where he studied the case of real~$\beta$ and~$\Omega_i$ and found~$\abs{\Omega_i}<1$, consistently  
with~\cite{Hawking:1998kw, Hawking:1999dp}.\footnote{For a concrete example, 
we generalize an argument in \cite{Hawking:1999dp}: consider a conformally-coupled scalar on $S^1_\beta \times S^3$, 
and expand in modes corresponding to the irreducible representations labelled $(\omega, j, j)$, 
corresponding to the quantum numbers $\{ E, \, J_L \equiv (J_2 + J_1)/2, \, J_R \equiv (J_2 - J_1)/2 \}$ 
introduced below \eqref{eq:4d_SCI_Background}. The partition function is 
\[
    Z_0 \, = \, \sum \e^{ - \beta ( \omega - \Omega_1 (m_L - m_R) - \Omega_2 (m_L + m_R))}  \,, 
\]
where the sum runs over~$j$, $m_L$, $m_R$, and~$\omega$, with the further condition coming from the scalar equation of motion that $\omega = 2j+1$, so that 
\[
\begin{split}
    Z_0 \, &= \, \sum_{j=0}^\infty \e^{-\beta ( 2j+1)} \sum_{m_L=-j}^j\sum_{m_R = -j}^j \e^{ \beta (\Omega_1+ \Omega_2) m_L + \beta (\Omega_1 - \Omega_2) m_R } \\
    \, &= \, \sum_{j=0}^\infty \e^{ - ( 2j+1) \beta }\frac{ \e^{- 2j \beta  \Omega_1 } \left( \e^{(2j+1)\beta  (\Omega_1-\Omega_2)} -1\right) 
    \left( \e^{(2j+1)\beta  (\Omega_1+\Omega_2)}-1\right)}{ \left( 
    \e^{\beta  (\Omega_1 - \Omega_2)}-1\right) \left(\e^{\beta  (\Omega_1+\Omega_2)}-1\right)} \, .
\end{split}
\]
The large $j$ behavior of the sum is governed by the following four exponential terms
\[
     \sim \, \e^{- 2 \beta (1-\Omega_1) j} - \e^{-2 \beta (1+\Omega_2) j} - \e^{- 2 \beta (1 - \Omega_2) j} + \e^{-2j\beta (1+\Omega_1)} 
\]
and convergence requires~\eqref{eq:KS_S3_PartitionFunction}. 
We notice that we cannot make the eigenvalues of $J_1$ and $J_2$ simultaneously large because of the geometry of $S^3$, which is a non-trivial fibration of $T^2$ (generated by $J_1$ and $J_2$) over the interval. 
At each endpoint of the interval one of the two circle fibres degenerates, so only one of the two angular momenta can become large. 
This leads to independent convergence conditions for each $\Omega_i$, 
as we found when deriving \eqref{eq:bKSW_sigma} and \eqref{eq:bKSW_tau}. In the trace defining the supersymmetric index, effectively only two of the four summands remain.
\label{footnote:HawkingReall}}

In these variables, the microscopic constraints~\eqref{eq:5d_AdSKN_Convergence_Conditions_1} 
and~\eqref{eq:5d_AdSKN_Convergence_Conditions_2} correspond to
\begin{equation}
\label{eq:KS_S3_Index}
    \Re\beta \, > \, 0 \, , \qquad \frac{\Re(\beta\Omega_1)}{\Re\beta} \, < \, 1 \, , \qquad \frac{\Re(\beta\Omega_2)}{\Re\beta} \, < \, 1 \, ,
\end{equation} 
which are too weak to guarantee the convergence of the sum over $\mc{H}_{S^3}$ in \eqref{eq:KS_ThermalPF}.
However, the superconformal index,  
which includes the insertion of $(-1)^F$ in the trace, 
can be expressed as the partition function 
with a certain background 
value of the $U(1)_R$ symmetry potential as follows (see~\eqref{eq:4d_PF_SCI_Fili}),  
\begin{equation}
\label{eq:4d_PF_SCI_Fili_2}
\begin{split}
    & Z(\beta, \Omega_1, \Omega_2, \Phi_R) \Big{\rvert}_{\beta(1 - 2\Phi_R + \Omega_1 + \Omega_2) \= 2\pi\ii (1 + 2n)} \, \\
    &\qquad \= \Tr_{\cH_{S^3}} \exp \bigl( - \beta \EnergyOp + \beta \Omega_1 J_1 + \beta \Omega_2 J_2 + \beta \Phi_R R \bigr) \Big{\rvert}_{\beta(1 - 2\Phi_R + \Omega_1 + \Omega_2) \= 2\pi\ii (1 + 2n)} \\
    &\qquad \= \Tr_{\cH_{\rm BPS}} \exp \left( \beta \left( \Omega_1 - 1 \right) \left( J_1 + \tfrac{1}{2}R \right) + \beta \left(\Omega_2 - 1\right) \left(J_2 + \tfrac{1}{2} R \right) \right) \, ,
\end{split}
\end{equation}
where in the last equality we have stressed that the sum runs only over the BPS subspace~\cite{Witten:1982df}. 
As already remarked around~\eqref{eq:5d_AdSKN_Convergence_Conditions_2}, the eigenvalues of~$J_{1,2} + \frac{1}{2}R$ are non-negative on $\cH_{\rm BPS}$, which means that convergence is guaranteed even when assuming only~\eqref{eq:KS_S3_Index}.

We noted below~\eqref{eq:KS_ThermalPF} that the KS criterion for the consistency of a conformal field theory 
on a complex $S^1\times S^3$ background matches the KSW criterion for the allowability of the asymptotically AdS$_5$ black hole. 
Here we see the same holographic correspondence in action in presence of fermions. 
As shown in this subsection, the conditions imposed by the supersymmetric index on the quantum states of the boundary superconformal field theory 
modify the allowability conditions of the bosonic theory. 
In the bulk, we have to impose supersymmetric boundary conditions on the gravitational theory dictated by the same supersymmetric index.
As shown in the previous subsection, 
the smooth saddle points of the resulting gravitational index give rise to the same modifications on the allowability conditions as the boundary theory.

\subsection{Expansion near the conformal boundary}

Next, we move to a neighborhood of the conformal boundary, where we define perturbatively a coordinate change from $(r,\BulkTheta)$ to $(z,\BdryTheta)$
\begin{equation}
\begin{split}
    \BulkTheta &\= \arcsin \left(\sin \BdryTheta \sqrt{\frac{\Xi_a}{\Delta_{\BdryTheta}}}\right) + z^2 \left( -\frac{ \sin \BdryTheta \cos \BdryTheta (a^2-b^2) \sqrt{\Xi_a \Xi_b } }{2 \Delta_{\BdryTheta}^2} \right) + o(z^3) \, , \\
    r &\= \frac{1}{z} \sqrt{\Delta_{\BdryTheta}} + z \left( - \frac{ \Delta_{\BdryTheta}^2 + 2 (\Xi_a a^2\sin^2\BdryTheta + \Xi_b b^2 \cos^2\BdryTheta ) }{4 \Delta_{\BdryTheta}^{3/2}} \right)  + o(z) \, ,
\end{split}
\end{equation}
such that the metric takes the form
\begin{equation}
\label{eq:Bdry_FG}
\begin{split}
    \rd s^2 &\= \frac{\rd z^2}{z^2} \left( 1 + o(z^2) \right) + \frac{1}{z^2} h_{ij}(x,z) \rd x^i \rd x^j \,, \\
    h(x,z) &\= g_{(0)} + z^2 \, g_{(2)} + o(z^2) \, , \\
    g_{(2)} 
    &= \frac{1}{2} \left[ \beta^2 \rd \tE^2 - \rd\BdryTheta^2 - \sin^2\BdryTheta \, ( \rd\phi - \ii \beta \Omega_1 \, \rd\tE )^2 - \cos^2\BdryTheta \, ( \rd\psi - \ii \beta \Omega_2 \, \rd\tE )^2 \right]
\end{split}
\end{equation}
where $g_{(0)}$ has line element $\rd s^2_{\rm bdry}$ as in \eqref{eq:4d_SCI_Background}. This represents a perturbative coordinate change to Fefferman--Graham coordinates: the leading term is \eqref{eq:AdS5_BdryMetric}, and one can check that $g_{(2)}$ is fixed by the Schouten tensor of $g_{(0)}$ as
\begin{equation}
    g_{(2),ij} = - \frac{1}{2} \left( R[g_{(0)}]_{ij} - \frac{1}{6} g_{(0),ij} R[g_{(0)}] \right) \, .
\end{equation}
As before, $z, \BdryTheta\in \R$, and we can focus on a three-dimensional complex line element on $\tilde{M}$ which in the real basis $\{ \rd \tE, \, \rd \phi , \, \rd\psi \}$ is represented by the matrix
\begin{equation}
    \tilde{h} \equiv \tilde{g}_{(0)} + z^2 \, \tilde{g}_{(2)} + o(z^2) \, , 
\end{equation}
where $\tilde{g}_{(0)}$ is \eqref{eq:gTilde0} and $\tilde{g}_{(2)}$ comes from $g_{(2)}$ in \eqref{eq:Bdry_FG}. We now apply the KSW criteria \eqref{eq:Allowability_RealPart} to $\tilde{h}$ to see whether the first subleading term in the Fefferman--Graham expansion near the conformal boundary imposes constraints on the parameters that are more stringent than those defining $\textsf{BKSW} = \textsf{micro}$.

The KSW condition with $q=0$ is $\Re \sqrt{\det \tilde{h}} > 0$, that is, 
\begin{equation}
    \frac{1}{\sin\BdryTheta \cos\BdryTheta}\Re \sqrt{\det \tilde{h}} \= \left( 1 - \frac{1}{4}z^2 + o(z^2) \right) \Re\beta \, > \, 0 \, .
\end{equation}
From the second rewriting we find that either $\Re\beta>0$ and $1 - \frac{1}{4}z^2 > o(z^2)$, or $\Re\beta < 0$ and $1 - \frac{1}{4}z^2 < o(z^2)$, but it's clear that the second possibility cannot happen as $z\to 0$, so we conclude that the KSW condition with $q=0$ applied to the metric including the first subleading correction near the conformal boundary is once again $\Re\beta > 0$, without additional constraints.

The KSW criterion with $q=1$, instead, requires us to check the positive-definiteness of $\Re W \equiv \Re \sqrt{\det \tilde{h}} \, \tilde{h}^{-1}$, which again we do by checking Sylvester's criterion. Here $\Re W$ has the form
\begin{equation}
\begin{split}
    \Re W &\= \left[ \Re W_{(0)} + z^2 \, \left( - \frac{3}{4} \Re W_{(0)} + D \right)  + o(z^2) \right] \, , \\
    D &\= \cos\BdryTheta\sin\BdryTheta \Re\beta \begin{pmatrix}
        0 & 0 & 0 \\ 0 & \frac{1}{\sin^2\BdryTheta} & 0 \\ 0 & 0 & \frac{1}{\cos^2\BdryTheta}
    \end{pmatrix}
\end{split}
\end{equation}
where $\Re W_{(0)}$ is \eqref{eq:ReW0}. Let $M^{(k)}$ denote the upper-left $k\times k$ block of $M$, then if $\det \Re W_{(0)}^{(k)} \neq 0$, the leading principal minor of order $k$ of $\Re W$ is 
\begin{equation}
    \det \Re W^{(k)} \= \det \Re W_{(0)}^{(k)} \left( 1 + z^2 \left( - \frac{3}{4}k + \sum_{i=1}^k ( \Re W_{(0)}^{(k)})^{-1}_{ii} D_{ii}^{(k)} \right) + o(z^2) \right) \, ,
\end{equation}
and we conclude, by an analogous argument to that just discussed, that the subleading term in the Fefferman--Graham expansion of the metric does not impose more constraints on $(a, r_+, r_\star)$ than the conformal boundary. That is, the analysis of the perturbation reduces to the analysis of the conditions on $\det \Re W_{(0)}^{(k)}$.\\
In particular, this is the case for the first leading principal minor, which is the product of $\left( 1 - \frac{3}{4}z^2 + o(z^2) \right)$ and the first principal minor of $\Re W_{(0)}$: we find once again that the condition is $\Re\beta>0$. \\
However, this doesn't have to be the case for the leading principal minors of order $k=2, 3$: it could happen that for a choice of parameters, $\det\Re W_{(0)}^{(k)}=0$, in which case the formula above cannot be applied, and the expansion of $\det \Re W^{(k)}$ generically starts at $z^2$ with a coefficient dependent on $\BdryTheta$ and $(a, r_+, r_\star)$.\\
Concretely, positivity of the minor of order $2$, $\det \Re W_{(0)}^{(2)}>0$, is equivalent to
\begin{equation}
    (\Re\beta)^2 - \sin^2\BdryTheta (\Re\beta  \Omega_1)^2 + \frac{z^2}{2} \left( 3 \sin^2\BdryTheta ( \Re \beta  \Omega_1)^2 - (\Re\beta)^2 \right) + o(z^2) \, > \, 0 \, ,
\end{equation}
but when the order-$1$ term vanishes, this inequality reduces to $z^2 \, \sin^2\BdryTheta (\Re \beta\Omega_1)^2 > o(z^2) $, which holds as $z\to 0$, so in this case even the boundary of the region allowed by $\textsf{BKSW}$ does not lead to additional constraints at subleading order. \\
Positivity of the minor of order $3$ is instead equivalent to
\begin{equation}
\begin{split}
    & \left((\Re\beta)^2 - \sin^2\BdryTheta (\Re\beta  \Omega_1)^2 - \cos^2\BdryTheta (\Re\beta  \Omega_2)^2 \right) \\
    & \qquad \qquad + \frac{z^2}{4} \left( 5 \sin^2\BdryTheta (\Re \beta \Omega_1)^2 + 5 \cos^2\BdryTheta (\Re\beta \Omega_2)^2 - (\Re\beta )^2 \right) + o(z^2) \, > \, 0 \, ,
\end{split}
\end{equation}
but again, when the term of order $1$ vanishes, this reduces to an inequality that always holds as $z\to 0$.

\smallskip

We conclude that the KSW criteria do not impose additional constraints on $(a, r_+, r_\star)$ when we move into the bulk from the boundary to first order in $z^2$.

\medskip

\subsection{KSW criterion in the bulk}

\begin{figure}[t]
  \centering
\begin{subfigure}{0.45\textwidth}
    \centering
        \begin{overpic}[width=\textwidth]{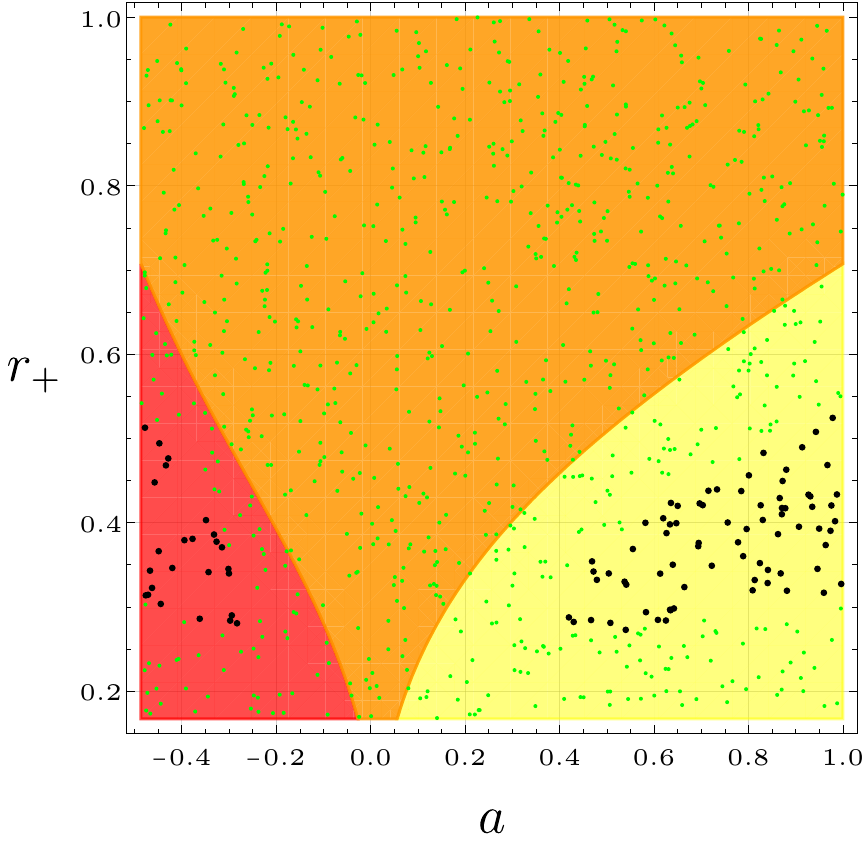}
                \put(9,194){\footnotesize $r_+$}
                \put(197,8){\footnotesize $a$}
        \end{overpic}
    \caption{$r = 4 r_+$}
    \label{fig:allowedregion_a}
  \end{subfigure}
  \hfill
  \begin{subfigure}{0.45\textwidth}
    \centering
        \begin{overpic}[width=\textwidth]{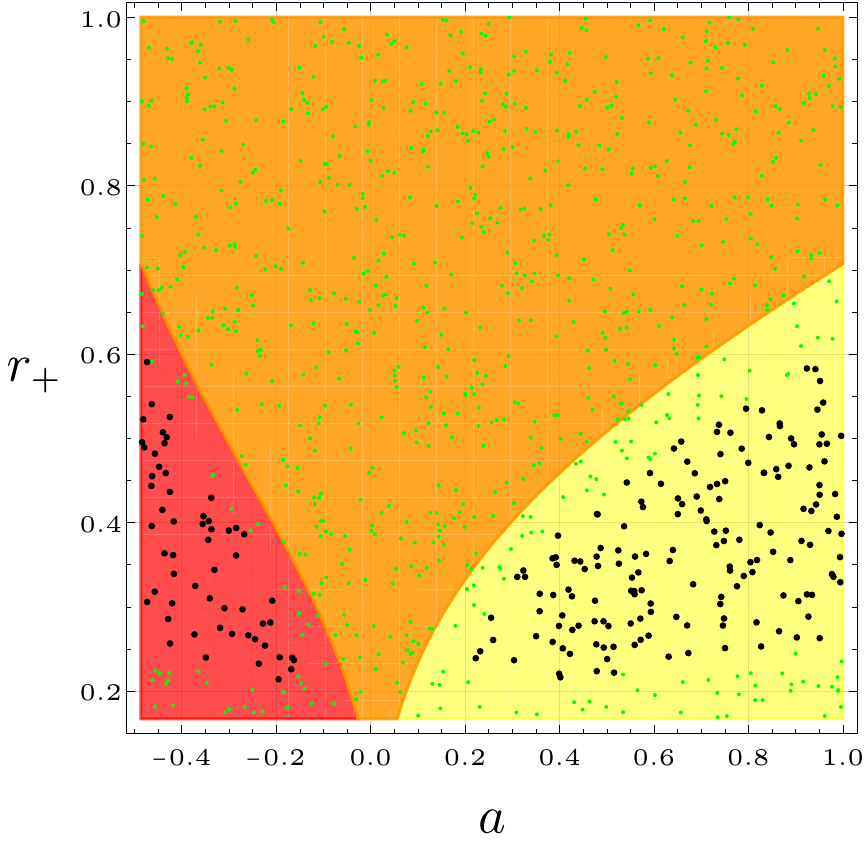}
                \put(9,194){\footnotesize $r_+$}
                \put(197,8){\footnotesize $a$}
        \end{overpic}
    \caption{$r = 5 r_+$}
    \label{fig:allowedregion_b}
  \end{subfigure}
\vskip\baselineskip
\begin{subfigure}{0.45\textwidth}
    \centering
        \begin{overpic}[width=\textwidth]{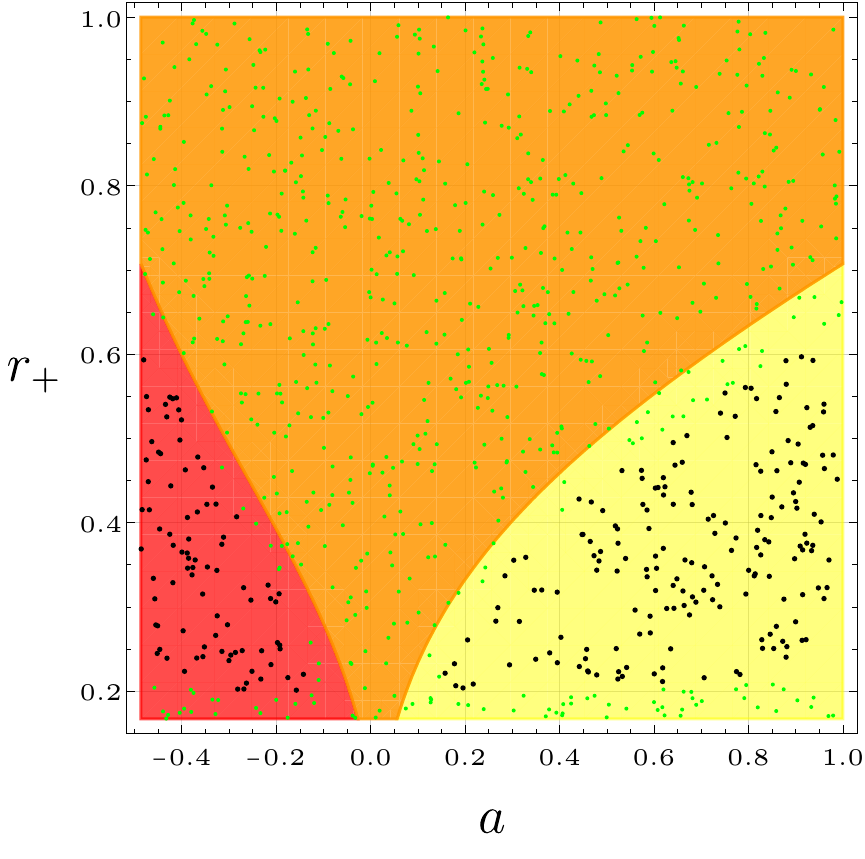}
                \put(9,194){\footnotesize $r_+$}
                \put(197,8){\footnotesize $a$}
        \end{overpic}
    \caption{$r = 6 r_+$}
    \label{fig:allowedregion_c}
  \end{subfigure}
  \hfill
  \begin{subfigure}{0.45\textwidth}
    \centering
    \begin{overpic}[width=\textwidth]{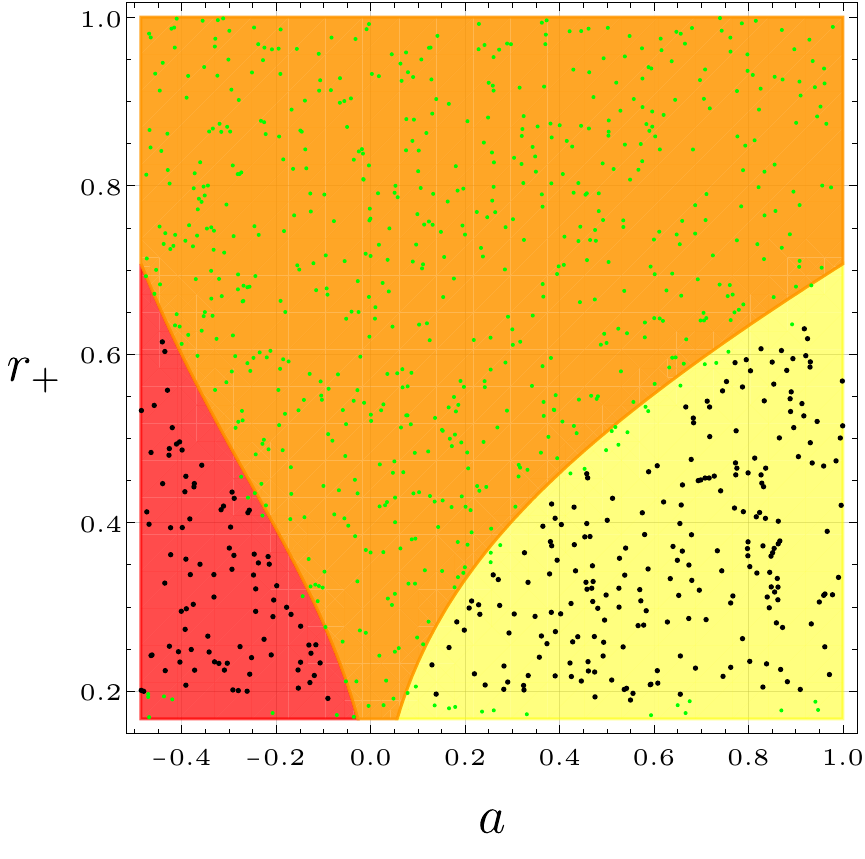}
                \put(9,194){\footnotesize $r_+$}
                \put(197,8){\footnotesize $a$}
    \end{overpic}
    \caption{$r = 8 r_+$}
    \label{fig:allowedregion_d}
  \end{subfigure}
\caption{\textit{Plot of microscopic constraints, geometric constraints and numerical evaluation of the KSW criteria.} The figure shows the regions $\textsf{KSW}_R$ defined in \eqref{eq:KSWR_def} for $R=nr_+$ and $n= 4, 5, 6, 8$. At fixed $r_\star = 1/6$, $a$ and $r_+$ are uniform-randomly picked within the geometric region, and the algorithm around \eqref{KSWeigenvalueEq} is run for all $\BulkTheta \in [0,\pi/2]$. A green dot denotes that the algorithm is passed, and thus the metric belongs to the family $\textsf{KSW}_{R}$ for the specific choice of $R$. In contrast, a black dot indicates that the KSW algorithm was not passed: there was a $\BulkTheta \in [0,\pi/2]$ for which \eqref{KSWeigenvalueEq} did not hold. We remark that the allowed region appears to shrink as $R$ increases, approaching the microscopic convergence region~\eqref{microregion} as $R \to \infty$.}
\label{fig:allowedregions}
\end{figure}
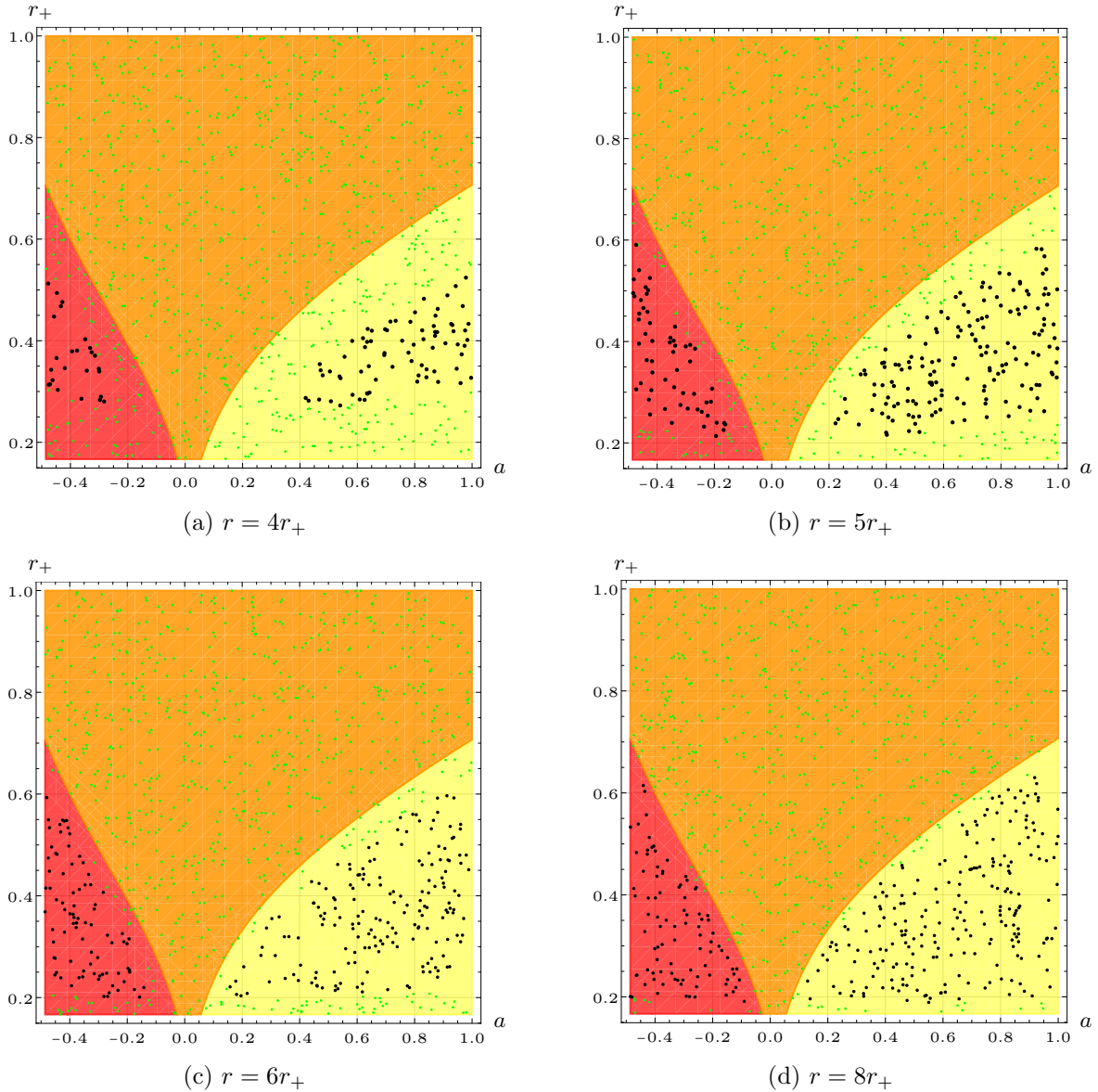

We now turn to the full bulk of the spacetime: the metric is \eqref{eq:AdS_KN_5d_Metric_App}, with $m$ fixed by \eqref{eq:AdS5_mSUSY} and the analytic continuation $t= - \ii \beta \tE$. In the coordinates $\{ \tE, r, \BulkTheta, \phi, \psi \}$, it can be represented as a $5 \times 5$ matrix $g_{\mu \nu}(r,\BulkTheta)\Big|_{(a,r_+,r_\star)}$, for which we must check the KSW criteria for all $r > r_+$ and $\BulkTheta \in [0,\pi/2]$. Because of the analysis at the conformal boundary leading to \eqref{trivialimplication1}, we can restrict ourselves to $(a, r_+, r_\star) \in \textsf{micro}$. 
The remaining question is whether there are such metrics for which $(a, r_+, r_\star) \notin \textsf{KSW}$, 
i.e.~whether KSW is more stringent than the microscopic criteria. 

We proceeded numerically by implementing the algorithm given around Equation~\eqref{KSWeigenvalueEq} 
by taking a representative sample of points $(r,\theta)$, 
and a representative sample of metrics 
determined by~$(a, r_+, r_\star)$ as follows. 
For $(a, r_+, r_\star)$, we chose $r_\star$ uniform-randomly 
in~$(0,\sqrt{3})$, then $a$ uniform-randomly in $((r_\star^2-1)/2,1)$ and finally $r_+$ uniform-randomly in $(r_\star, 5 r_\star)$. In this way 
we collected 1000 3-tuples $(a, r_+, r_\star)$, which we then sieved by the microscopic convergence criteria in~\eqref{microregion}, leaving us with 894 3-tuples. For each of these, we sampled 100 pairs $(r,\BulkTheta)$ 
by choosing $r$ uniform-randomly in $(r_+, 5 r_+)$ and $\BulkTheta$ uniform-randomly in~$[0,\pi/2]$.  
In this sample, we did not find any violations of the KSW criteria. 
Combining this with the analytic proof~$\textsf{KSW} \subseteq \textsf{micro}$ in the previous section, 
these numerical experiments build confidence towards the proposition that
\begin{equation}
	\textsf{KSW} \= \textsf{micro} \text{ for the supersymmetric index} \,.
\end{equation}

Although, of course, numerics cannot conclusively prove the above equivalence, it can be used effectively as an exclusion criterion. 
In other words, if we had found a $g \in \textsf{micro}$ such that KSW is \textit{not} satisfied at any point in the bulk, 
then we would be able to conclude $\textsf{KSW} \neq \textsf{micro}$ and is only a proper subset. 
However, we did not find such instances.

Finally, since we have presented evidence that the KSW criteria on $(a,r_+,r_\star)$ are most stringent near the conformal boundary, 
one might wonder how they behave as we increase~$r$ from~$r_+$ to~$\infty$. 
We were unable to find a statement of monotonicity of any simple analytic function of 
the metric, such as the left-hand side of~\eqref{KSWeigenvalueEq}. 
Hence we examined this question numerically, but 
also here we only began to scratch the surface: 
we picked $r_\star = 1/6$, then $a$ uniform-randomly in $((r_\star^2-1)/2,1)$ and $r_+$ uniform-randomly in $(r_\star,1)$. This time we did not filter these $(a,r_+,r_\star)$ 
further by the microscopic convergence criteria, but instead kept them more generally in the geometric regime~\eqref{eq:Geometric_Constraint_Degeneracy_v2}--\eqref{eq:Geometric_Constraint_rPlus_rStar}. 
Then, we considered $g_{\mu \nu}(r,\BulkTheta)$ on the radial slice $r = n r_+$ for $n = 4,5,6,8$,\footnote{It is not the induced metric on the slice, 
but simply the $5 \times 5$ matrix $g_{\mu \nu}(n r_+,\BulkTheta)$. For each $n$ we chose 1000 tuples $(a,r_+,r_\star)$.} 
and examined the KSW algorithm around Equation~\eqref{KSWeigenvalueEq} for all $\BulkTheta \in [0,\pi/2]$. We denote the KSW constraints on $(a,r_+,r_\star)$ from the points on the slice at a constant $r = R$ only as
\begin{equation}
\label{eq:KSWR_def}
	\textsf{KSW}_R \, \equiv \, \{ (a,r_+,r_\star) \ \big| \ \eqref{eq:Allowability_RealPart} \text{ holds for } g_{r=R} \text{ on } M_{r=R} \} \,,
\end{equation}
which is of course dependent on our specific choice of coordinates $\{ \tE,r,\BulkTheta,\phi,\psi \}$ on $M$. Our earlier analytic analysis showed that $\textsf{KSW}_{R\to \infty} = \textsf{BKSW} = \textsf{micro}$. 
The numerical procedure just outlined further suggests the \textit{monotonic} behavior
\begin{equation} 
\label{KSWmono}
	\textsf{KSW}_{R_2} \, \subset \, \textsf{KSW}_{R_1} \qquad \text{if } R_2 \, > \, R_1 \,.
\end{equation}
We summarize our evidence for this proposition in Figure~\ref{fig:allowedregions}. We leave it to future work to prove/disprove~\eqref{KSWmono} and to understand its possible generality.

\section*{Acknowledgements}
We are grateful to Marco Ambrosini and Edward Witten for helpful communications. PBG is supported by the SNSF Ambizione grant PZ00P2$\_$208666. S.M.~acknowledges the support of the STFC grants ST/T000759/1,  ST/X000753/1 during the course of this work.

\appendix

\section{A practical algorithm for implementing the KSW criterion} 
\label{appA}

In this appendix we prove the algorithm stated around Eq.~\eqref{KSWeigenvalueEq}, which is a characterization of KSW-allowability. 
This algorithm is slightly different from the statement of the characterization given  
in~\cite{Kontsevich:2021dmb,Witten:2021nzp} (as reviewed around  Eq.~\eqref{eq:QuadraticForm_Diagonal}),
and brings practical advantages in its implementation. 
In~\eqref{eq:QuadraticForm_Diagonal} the~$\lambda_i$ are the diagonal elements of $g_p$ in a \textit{certain} real coordinate basis in which $g_p$ is diagonal. 
Such a basis is guaranteed to exist if the~$q=1$ condition in~\eqref{eq:Allowability_RealPart} is satisfied.
In order to perform computations, one must first identify such a basis. 
In contrast, in the algorithm containing  Eq.~\eqref{KSWeigenvalueEq}, the $\mu_i(g_p A_p)$ are the eigenvalues of $g_p A_p$ where $g_p$ is expressed in \textit{any} real coordinate basis. 
The step of first putting~$g_p$ in diagonal form by a suitable real coordinate transformation is hence eliminated (however, one must still first check that $A_p$ is positive definite, but this can also be done in any basis). Most of our argument closely follows \cite{Witten:2021nzp}.

\smallskip

The KSW criteria~\eqref{eq:Allowability_RealPart} must hold at all points $p \in M$ and for all \textit{real-valued} $q$-forms~$F$. 
These conditions are independent of the \textit{real} basis in which we choose to view $g = (g_{\mu \nu})$ as a matrix with inverse $g^{-1} = (g^{\mu \nu})$ and $F = (F_{\mu_1 \mu_2 \cdots \mu_q})$ as a totally antisymmetric object with $q$ indices. 
From here on we will work in a particular real coordinate basis of our choosing, and suppress the $p$-dependence. 

The $q = 0$ condition tells us we should be able to make one sign choice for $\sqrt{\det g}$ over the whole manifold such that $\text{Re} \sqrt{\det g} > 0$. 
It means that $\det g$ may never become negative real on $M$.

The $q=1$ condition tells us that $W = \sqrt{\det g} \, g^{-1}$ can be written as $W = A + \ii B$ where $A$ and $B$ are real symmetric matrices and $A$ is positive definite. 
This means that we can 
write $A = V^\top V$ for a real invertible matrix $V$. 
The matrix $(V^{-1})^\top \, B \, V^{-1}$ is then real symmetric, and hence can be orthogonally diagonalized by $O$ as $O^\top (V^{-1})^\top B V^{-1} O = \tilde{D}$. 
With $P \equiv V^{-1} O$, we then have 
\begin{equation} \label{eq:PPtrels}
	 P^\top A \, P \= \mathds{1} \,, \quad  P^\top B \, P \= \widetilde D  \quad \Rightarrow \quad  
     P^\top W \, P \= \mathds{1} + \ii \, \widetilde D \; \equiv \; D \,.
\end{equation}
Note that $P$ is not necessarily orthogonal because $V$ is not, and hence~$W$ is diagonalized by congruence -- as a quadratic form is -- and not by a similarity transformation -- as a linear operator would be.  
Nevertheless~$P$ is a real invertible matrix, which is useful below. 
Further, the elements of~$D = \text{diag}(d_j) = \text{diag}(1+ \ii \tilde{d}_j)$  
are \textit{not}  the eigenvalues of $g$ (we return to the meaning of~$d_j$ below). We have $S^\top g S = \sqrt{\det g} \, D^{-1}$ with $S = (P^{-1})^\top$, so that the $\lambda_i$ in \eqref{eq:Allowability_Eigenvalues} are equal to $\sqrt{\det g} \, d_i^{-1}$.

Then, we have 
\begin{equation}
	g^{-1} \= \frac{1}{\sqrt{\det g}} \, Q^\top D \, Q \,,
\end{equation}
where $Q \equiv P^{-1}$ is real. In components,
\begin{equation}
	g^{ij} \= \frac{1}{\sqrt{\det g}} \sum_k Q^{ki} Q^{kj} d_k
\end{equation}
and defining $\tilde{F}^{k_1 k_2 \cdots k_q} \equiv Q^{k_1 i_1} Q^{k_2 i_2} \cdots Q^{k_q i_q} F_{i_1 i_2 \cdots i_q}$, the condition \eqref{eq:Allowability_RealPart} becomes
\begin{equation}
	\sum_{\boldsymbol{k}} (\tilde{F}^{k_1 \cdots k_q})^2 \, \Re\left( \sqrt{\det g} \frac{d_{k_1}}{\sqrt{\det g}} \cdots \frac{d_{k_q}}{\sqrt{\det g}} \right) \; > \; 0
\end{equation}
for all $F$ (equivalently, all~$\tilde{F}$) and $q \in \{ 0,1,\cdots,D \}$. In other words, 
\begin{equation} 
\label{dicondition}
	\Re \left( \sqrt{\det g} \, \prod_{i \in S} \frac{d_i}{\sqrt{\det g}} \right) \; > \; 0
\end{equation}
for all subsets $S \subseteq \{ 1,2,\cdots,D \}$.

\smallskip

Returning to the $d_i$: we now show that they are proportional to the eigenvalues of $(gA)^{-1}$. To see this, note that the $d_i$ are the solutions to the characteristic equation
\begin{equation}
	\det \left( D - d \, \mathds{1} \right) \= 0 \,.
\end{equation}
Now, the following sequence of manipulations, 
\be
\begin{split}
    \det \left( D - d \, \mathds{1} \right) 
    &\= \det \left(P^\top W \, P - d \, P^\top A \, P \right)  \\
    &\= \det \left(P^\top A \, \left( A^{-1} \, W  - d \, \mathds{1} \right) P \right)  \\
    &\= \det \bigl(P^\top A \bigr)\, \det\left( A^{-1} \, W  - d \, \mathds{1} \right) \, \det \left( P \right)  \\
    &\=  \det\left( A^{-1} \, W  - d \, \mathds{1} \right)  \,,
\end{split}
\ee
where we have used~$P^\top A \, P = \mathds{1}$ \eqref{eq:PPtrels} in the first and fourth equalities,
shows that~$d_i$ are the solutions to the characteristic equation
\be
    \det \Bigl( \sqrt{\det g} \, A^{-1} g^{-1} - d \, \mathds{1} \Bigr) \,.
\ee
In other words, recalling the notation $\mu_i ( \, \cdot \, )$ for ``eigenvalue of'', 
\begin{equation}
	d_i 
    \= \sqrt{\det g} \; \mu_i \bigl((g A)^{-1} \bigr) \,.
\end{equation}
Therefore, the positivity conditions \eqref{dicondition} can be written as (recall $\det A > 0$):
\begin{equation} 
\label{gAcondition}
	\Re \Bigl( \sqrt{\det g A} \, \prod_{i \in S} \mu_i \bigl( (gA)^{-1} \bigr) \Bigr) \, > \, 0
\end{equation}
for all subsets $S \subseteq \{ 1,2,\cdots,D \}$. This is equivalent to
\begin{equation} 
\label{argcondition2}
	\sum_{i=1}^D \abs{\Arg \mu_i( gA )} \, < \, \pi \,.
\end{equation}

It is clear that the above conclusion does not change under real rescalings and permutations of the $d_i$.

\bigskip
\bigskip

\bibliographystyle{JHEP}
{\small
\bibliography{Bib_BH}}

\end{document}